# Nanoscale Dynamic Readout of a Chemical Redox Process Using Radicals Coupled with Nitrogen-Vacancy Centers in Nanodiamonds

*Jan Barton[1,2,#], Michal Gulka[3,4,5,#], Jan Tarabek[1], Yuliya Mindarava[6], Zhenyu Wang[7], Jiri Schimer[1], Helena Raabova[1], Jan Bednar[8,9], Martin B. Plenio[7], Fedor Jelezko[6], Milos Nesladek[3,4,5], Petr Cigler[1,*]*

[1] Institute of Organic Chemistry and Biochemistry of the Czech Academy of Sciences, Flemingovo namesti 2, 166 10 Prague, Czechia

[2] Department of Inorganic Chemistry, Faculty of Science, Charles University, Hlavova 2030, 128 40 Prague 2, Czechia

[3] Institute for Materials Research (IMO), Hasselt University, Wetenschapspark 1, B-3590 Diepenbeek, Belgium

[4] Department of Biomedical Technology, Faculty of Biomedical Engineering, Czech Technical University in Prague, Sitna sq. 3105, 27201 Kladno, Czechia

[5] IMOMEC division, IMEC, Wetenschapspark 1, B-3590 Diepenbeek, Belgium.

[6] Institute for Quantum Optics and IQST, Ulm University, Albert-Einstein-Allee 11, D-89081 Ulm, Germany

[7] Institute of Theoretical Physics and IQST, Ulm University, Albert-Einstein-Allee 11, D-89081 Ulm, Germany

[8] Institute for Advanced Biosciences, UMR 5309, Allée des Alpes, 38700 la Tronche, France

[9] Institute of Biology and Medical Genetics, 1st Faculty of Medicine, Charles University, Albertov 4, 128 00 Prague, Czechia

[#] Equal contribution.

* Corresponding Author: Petr Cigler (cigler@uochb.cas.cz).

# ABSTRACT

Biocompatible nanoscale probes for sensitive detection of paramagnetic species and molecules associated with their (bio)chemical transformations would provide a desirable tool for a better understanding of cellular redox processes. Here, we describe an analytical tool based on quantum sensing techniques. We magnetically coupled negatively charged nitrogen-vacancy (NV) centers in nanodiamonds (NDs) with nitroxide radicals present in a bioinert polymer coating of the NDs. We demonstrated that the $T_1$ spin relaxation time of NV centers is very sensitive to the number of nitroxide radicals, with a resolution down to ~10 spins per ND (detection of approximately $10^{-23}$ mol in a localized volume). The detection is based on $T_1$ shortening upon the radical attachment and we propose a theoretical model describing this phenomenon. We further show this colloidally stable, water-soluble system can be used dynamically for spatiotemporal readout of a redox chemical process (oxidation of ascorbic acid) occurring near the ND surface in an aqueous environment under ambient conditions.

Recently, quantum optical approaches for localized detection of molecular and cellular processes at nanoscale have emerged, including optically detected magnetic resonance using negatively charged nitrogen-vacancy (NV) centers in diamond.[1,2] Quantum sensing with NV centers is based on the interaction of the NV electronic spin with its surroundings.[3] The NV acts as a room-temperature qubit that can be polarized with high efficiency and readout optically. Many innovative applications exploiting NV centers have been developed in the past decade.[4–12] For example, engineered covalent binding of a spin label on the surface of a single crystal diamond enabled detection of the attached radicals by means of Hahn echo and double electron-electron resonance (DEER).[13] For localized sensing inside living cells, nanodiamond particles (NDs) are particularly well suited and have become one of the most studied systems for nanoscale quantum sensing and metrology,[4,12] thanks to their biocompatibility and the favorable spectral properties of NV centers,[14]

Sensing with the NV centers can be achieved by $T_1$ relaxometry measurements, as the $T_1$ longitudinal spin relaxation time strongly depends on the fluctuations of the surrounding spin density, even more than $T_2$ relaxation time.[15] Additionally, $T_1$ relaxometry can be used as an all-optical technique without the need for microwave (MW) excitation,[16] which is commonly used in NV quantum-sensing protocols. MW-free measurements are advantageous for detection in biological systems as the heating effects and other interferences with the biological sample due to MW irradiation are avoided. The readout time for $T_1$ relaxometry is determined by the number of repetitions of the measurement sequence, needed to acquire a sufficient number of detected photons. Depending on the NV concentration in the optical focus, the execution of measurements can range from milliseconds for NV ensembles to about a minute for a single NV.[16] Detection of magnetic noise using relaxometry has been demonstrated for a range of paramagnetic species, including ferritin,[17–20] gadolinium,[16,21–25] manganese ions,[20,26] and superparamagnetic nanoparticles[27] with the possibility of ND-based sensing inside living cells.[28] This method has high sensitivity, as evidenced by the detection of a few ferritin molecules[17] or approximately 5 gadolinium spins per Hz$^{1/2}$ in a model biological environment.[22] Recently, relaxometry has been shown useful for pH sensing,[29] and the $T_1$ protocol has been enhanced to take into account ionization and recharging of the NV centers.[30] Additionally, a similar approach for dynamic quantum sensing of paramagnetic species based on spin-dependent emission of photoluminescence was recently established.[31] Specific detection of chemical events using $T_1$ relaxometry is also possible but requires a chemical[16] or mechano-chemical transducer.[32,33]

Chemical reactions of great scientific interest are cellular oxidation-reduction (redox) processes. Redox reactions play a vital role in maintaining metabolic and other functions in living cells.[34,35] These include energy production, cellular respiration, proliferation, differentiation, and apoptosis.[34] Localized monitoring of redox reactions is essential for understanding the cellular processes responsible for cell viability.[36] Detection of free radicals using fluorescent NDs has previously been proposed[37] but has not yet been demonstrated. Despite the high sensitivity of quantum optical approaches, their use to detect antioxidants has also not been established.

In this work, we present an analytical tool that enables sensitive characterization of radical species and spatiotemporal detection of redox reactions in an aqueous environment under ambient conditions. This tool is based on quantum sensing techniques, using so-called solid state qubits, and utilizes the electron spin of NV centers in NDs. Specifically, we developed a polymer architecture that allowed us to covalently attach nitroxide radicals near the surface of NDs containing NV centers (Scheme 1). We used a specific quantum sensing methodology employing the $T_1$ relaxation time measurements, where the laser beam is focused on a single ND particle in an aqueous environment. This methodology enables the quantitative determination of the concentration of nitroxide radicals bound on the ND surface with high sensitivity. Our technique is based on the shortening of $T_1$ relaxation time of NV centers when they are exposed to a fluctuating magnetic field of the radicals. We calibrated the $T_1$ relaxometric measurements with a bulk electron paramagnetic resonance (EPR) method, and theoretically modeled the results based on detection of the quantum magnetic noise induced by the presence of nitroxide electron spins on the ND surface. Finally, we demonstrated the use of this ND system as a highly sensitive relaxometric nanosensor with dynamic readout. We monitored reaction kinetics of the nitroxides with a major natural antioxidant, ascorbic acid, by measuring the $T_1$ time recovery at the single-particle level.

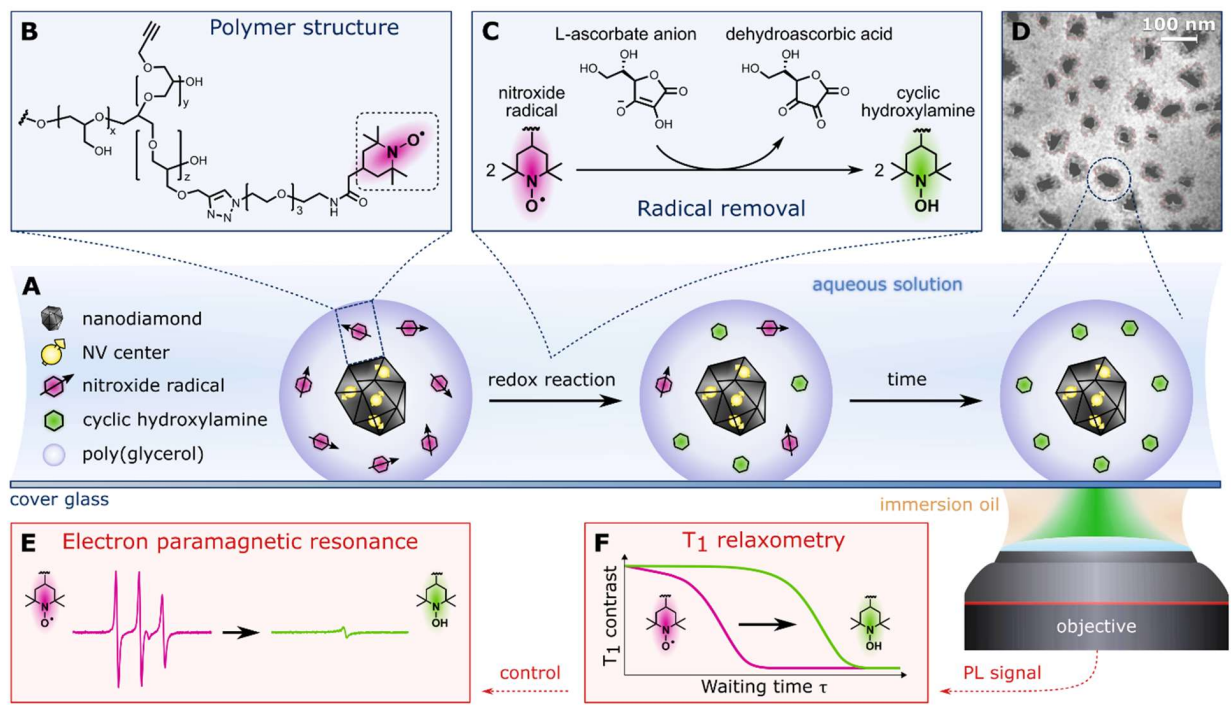

**Scheme 1:** Structure of ND nanosensor and experimental methods used for sensing of the radical presence. A) Schematic of the ND coated with a polymer shell containing spin probes coupled magnetically with NV centers. The radicals in nanosensor are gradually reduced by a redox reaction until they are all removed. B) Structure of modified poly(glycerol) shell bearing TEMPO-based nitroxide spin probes. The structure of the polymer is shown schematically; many branching points are present in the final dendritic structure of poly(glycerol). C) Redox reaction of the nitroxide radical with L-ascorbate, providing a diamagnetic cyclic hydroxylamine and dehydroascorbic acid. D) Cryo-TEM micrograph obtained from an aqueous solution of ND5. The dark objects are NDs; the contours of the polymer shell (red) were detected automatically. The average thickness of the polymer layer was 17.0 ± 8.8 nm. For larger micrographs, see Figure S4 in Supporting Information. E, F) Processes used for radical quantification and detection of ascorbate: EPR in bulk solution and optical $T_1$ relaxometry of NV centers at the single-particle level.

## RESULTS AND DISCUSSION

**Nanosensor design, preparation, and structure.** Our design of a relaxometric nanosensor is based on the central hypothesis of a highly sensitive magnetic coupling between NV centers in NDs and electron spins of radicals present in proximity to the NDs. Previous reports indicate a strong spin interaction between NV centers in bulk diamond and the radicals TEMPO[13] and 2,2-diphenyl-1-picrylhydrazyl[38] immobilized on the diamond surface. These findings suggest that NV centers can be used as a probe to sense molecular electron and nuclear spins present in nanometric proximity to the diamond surface. For example, EPR detection of an individual, nitroxide-labeled DNA duplex attached to NDs was recently achieved using a single NV center[39] and sensitive detection of ~$10^{-19}$ mol radicals was achieved using quantum probe relaxation microscopy.[40] However, measurements presented to date have focused on the detection of an equilibrium state.

In other words, no chemical processes have been monitored using NV centers coupled with a radical spin probe.

To achieve the goal of dynamic monitoring of chemical redox processes, we developed a polymer architecture that anchors nitroxide radicals on the ND surface and colloidally stabilizes the NDs (Scheme 1). We found that magnetic coupling between nitroxide radicals and NV centers can be utilized for selective relaxometric optical sensing of ascorbate in an aqueous environment with diffraction-limited resolution. We used TEMPO-based nitroxides, which are shelf-stable compounds that can be stored under ambient conditions. $T_1$ relaxometry measurements on a single ND particle with a few NV centers required ~1 min readout, which is negligible compared to the half-life of TEMPO-based nitroxides in an aqueous environment.

Nitroxides have natural chemical reactivity as selective oxidants. Specific redox reactions between target reducing species and nitroxides thus in principle would enable direct relaxometric detection of the reducing analytes. This is an important advantage compared to paramagnetic $Gd^{3+}$-complexes and magnetic nanoparticles, which must be embedded in a structurally responsive chemical transducer present on NDs to ensure distance control for $T_1$ or $T_2$ sensing.[16,32,33] Additionally, the smaller volume occupied by nitroxide probes compared to bulky $Gd^{3+}$-complexes and magnetic nanoparticles allows for more flexible tuning of the nitroxide concentration range embedded in a polymer matrix.

The hydrophobicity of TEMPO-based nitroxides requires efficient colloidal stabilization of the nanoparticles intended for biologically relevant sensing.[41] Previous studies simply modified the diamond surface using silica.[13,42,43] However, silica coating without additional polymeric coating does not sufficiently stabilize NDs in a biological environment,[44,45] especially in combination with the hydrophobic molecules[46] such as TEMPO. Our approach is based on i) increasing the solubility of TEMPO by attaching a hydrophilic linker and ii) creating a branched, structurally dense protecting layer of alkynated hydrogel poly(glycerol)[47,48] that can accommodate TEMPO on the ND surface (Scheme 1B). Details on synthesis of the modified cyclic hydroxylamine **3** (precursor of TEMPO radical) are described in Supporting Information. The polymer coating and modification of NDs by TEMPO radical are summarized in Scheme 2 and described in Materials and methods.

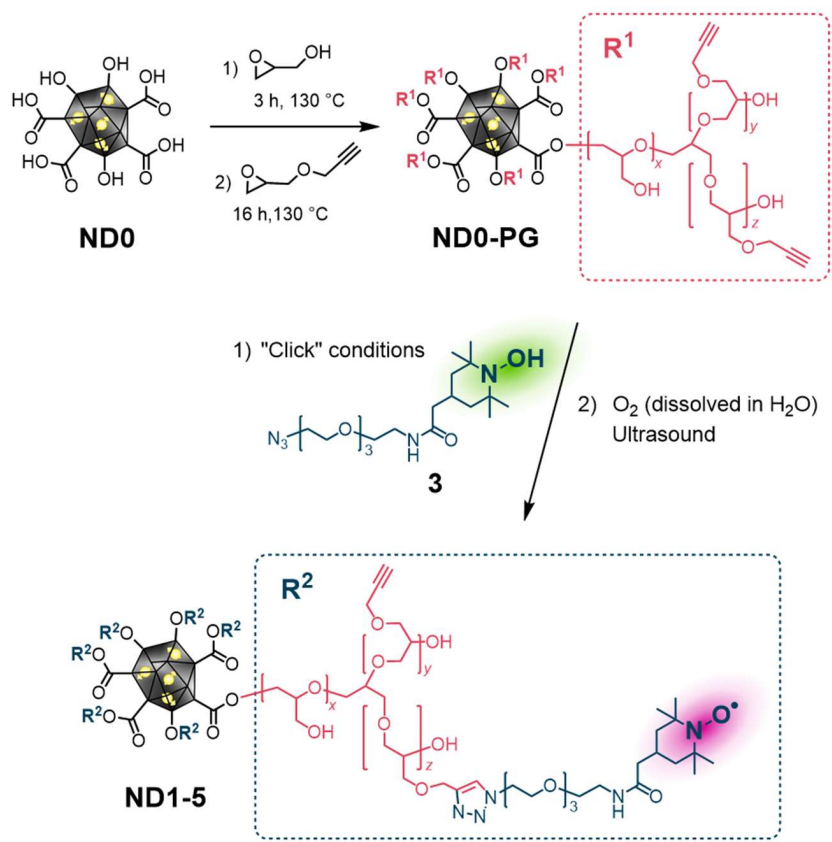

**Scheme 2:** Surface modification of NDs. The oxidized fluorescent **ND0** is first coated with poly(glycerol) and modified by alkyne moieties, providing **ND0-PG** reactive intermediate. The cyclic hydroxylamine is introduced using click chemistry and converted by air oxidation in ultrasound to TEMPO-type nitroxide radicals present on particles **ND1-5**.

We prepared NDs containing 64 weight % poly(glycerol) (Figure S1 in Supporting Information) modified with alkyne moieties. Using different concentrations of nitroxide precursor, we obtained a series of five NDs with gradually increasing loads of nitroxide radicals (**ND1-5**). Covalent attachment of the nitroxide using click chemistry was validated by infrared spectroscopy (decrease in the alkyne band and appearance of an amide I band present in the TEMPO molecule; Figure S2 in Supporting Information). We analyzed aqueous solutions of NDs using dynamic light scattering (DLS). All prepared particles showed excellent colloidal stability without a shift in size distribution even at the highest nitroxide load (**ND5**) (Table 1 and Figure S3 in Supporting Information).

**Table 1:** Characterization of the samples: nitroxide radical loads, the corresponding $T_1$ relaxation times, and hydrodynamic diameters in water determined by DLS. N.D. – not determined.

| Sample | Hydrodynamic diameter (nm) | Stable nitroxide radicals: average per particle | $T_1$ relaxation time (µs) |
|---|---|---|---|
| "naked" **ND0** (no polymer) | 81 | 0 | N.D. |
| **ND0-PG** | 112 | 0 | N.D. |
| **ND1** | 108 | 0 | 99.0 |
| **ND2** | 109 | 10 | 78.6 |
| **ND3** | 108 | 46 | 70.6 |
| **ND4** | 113 | 88 | 45.0 |
| **ND5** | 112 | 134 | 22.0 |
| **ND5** after reduction | N.D. | 0 | 100 |

Next, we studied the properties of the polymer layer using cryo-TEM, which enables the observation of macromolecules and nanoparticles in their native solution state. We analyzed the actual thickness of our polymer shells on NDs, which is essential for calculating the concentration of free spins around NDs. The polyglycerol layer appeared in cryo-TEM as a geometrically irregular, 17.0 ± 8.8 nm thick structure enveloping the NDs (Scheme 1D; Figure S4 in Supporting Information). The presence of the poly(glycerol) caused a natural steric separation of the particles in aqueous solution, which corresponds to the colloidal stability observed by DLS (Figure S3). Interestingly, the polymer layer was also detectable using TEM after sample staining with tungstic acid (Figure S5 in Supporting Information). The structural collapse of the layer caused by dehydration *in vacuo* led to the contraction of the polymer thickness to 5.7 ± 1.6 nm, corresponding to a reduction in the original volume occupied by the polymer to 17%. The observed collapse demonstrates the crucial role of water in the stabilization of the polymer structure. It further suggests the magnitude of errors that could be caused by artifacts from inappropriate TEM characterization and highlights the general importance of using cryo-TEM for characterization of polymer structures on nanoparticles.

**Radicals on NDs: quantification using EPR.** To determine the sensitivity of the detection scheme and the magnetic coupling between nitroxides and NV centers in our system, we first estimated the number of radicals covalently attached to the NDs. We used a complementary set of methods: EPR, cryo-TEM, TEM image analysis, nanoparticle tracking analysis (NTA) and thermogravimetry (TGA).

First, we quantified the concentration of radicals in all the samples using EPR executed on macroscopic volumes at low microwave power to avoid saturation effects. We deconvoluted the EPR spectra into omnipresent "sub-surface" radicals (overlapping signals of P1 centers and non-saturated dangling carbon bonds carrying unpaired electron spin[49]) and nitroxide radicals (see Supporting Information and Figure S6 for details) and quantified the absolute amount of each type of radical in the samples (Table S1 in Supporting Information). The ratios of nitroxide/"sub-surface" radicals ranged from 0 (**ND1**) to 2.75 (**ND5**), reaching a maximum concentration of nitroxide radicals of $4.6 \cdot 10^{18}$ $g^{-1}$.

To determine the radical concentration in the polymer layer and the number of radicals attached to an average ND particle, we combined the EPR data with other methods. From TEM image analysis, we obtained the volume-weighted distribution of ND core sizes in spherical approximation (without the polymer layer; Figure S7A in Supporting Information). Cryo-TEM enabled us to estimate the average native thickness of the polymer layer ($17.0 \pm 8.8$ nm), which was independent of particle size (Scheme 1D and Figure S4 in Supporting Information). In our geometric model, we formally added this polymer thickness to each individual particle identified by TEM image analysis. The total volume of polymer present in the sample was obtained as a sum of volume contributions from all individual particles. The concentration of nitroxide radicals obtained from EPR was recalculated to this volume for each ND sample (Table 1). Based on this number and known concentrations of both polymer and diamond in solutions analyzed thermogravimetrically (Figure S1 in Supporting Information), we estimated a number-weighted average nitroxide concentration per particle for each sample (Table 1). The highest average value was 134 nitroxide radicals per particle for **ND5**. Interestingly, the natural polydispersity of the samples results in a non-linear, approximately one-order-of-magnitude span of individual loads from ~70 to ~500 nitroxide radicals per particle for 10 to 60 nm core particles (Figure S7C in Supporting Information). Finally, these data showed a non-linear dependence of conjugation yield on the concentration of the radical precursor (Figure S7B in Supporting Information).

**Interaction of nitroxide radicals with NV centers: optical relaxometry at the single-particle level.** To quantify the influence of the radicals attached to NDs on the NV spin longitudinal $T_1$ relaxation time, we used a confocal microscope equipped for $T_1$ spin relaxometry. NDs in aqueous solution were deposited onto a cover glass, and the system was sealed to avoid evaporation, which could cause a collapse of the polymer layer. We characterized single particles using a pulsed $T_1$ sequence (Figure 1A) consisting of repetitive laser pulses without application of MW (see Materials and methods for more details), probing spin relaxation from the $m_s = 0$ spin state to the thermally mixed state.[21] We carried out the relaxometry measurements on randomly selected single particles (see Figure S8 in Supporting Information for histograms) and we averaged the resulting $T_1$ constants for each sample. The $T_1$ decreased with an increasing amount of nitroxide radicals attached to the polymer shell (Figure 1B). For the sample with maximal radical load (**ND5**), the average $T_1$ time reached a minimum of 24 μs. The mean $T_1$ time of **ND2** (with 10 radicals per particle on average, corresponding to $1.7 \cdot 10^{-23}$ mol) was approximately 15% lower compared to that of **ND1**, for which no radicals were detected by EPR. These results clearly demonstrate the exceptional potency of NV relaxometry to directly sense the presence of only a few radicals on the ND surface. However, even though the particles in this study were selected randomly, a few of the brighter NDs were skipped. These particles exhibited NV charge-switching as expected for high NV-density NDs.[50–52] Better uniformity of NDs in terms of particle size and NV content[53–55] would help reduce the dispersion of the observed $T_1$ times. A high $T_1$ spread, especially for ND samples with low or no radical load (see the error bars in Figure 1B), is expected. Some NVs can experience a shorter $T_1$ time due to factors not relevant to the presence of nitroxide radicals, such as interactions with a layer of electron spins located nearby the ND surface.

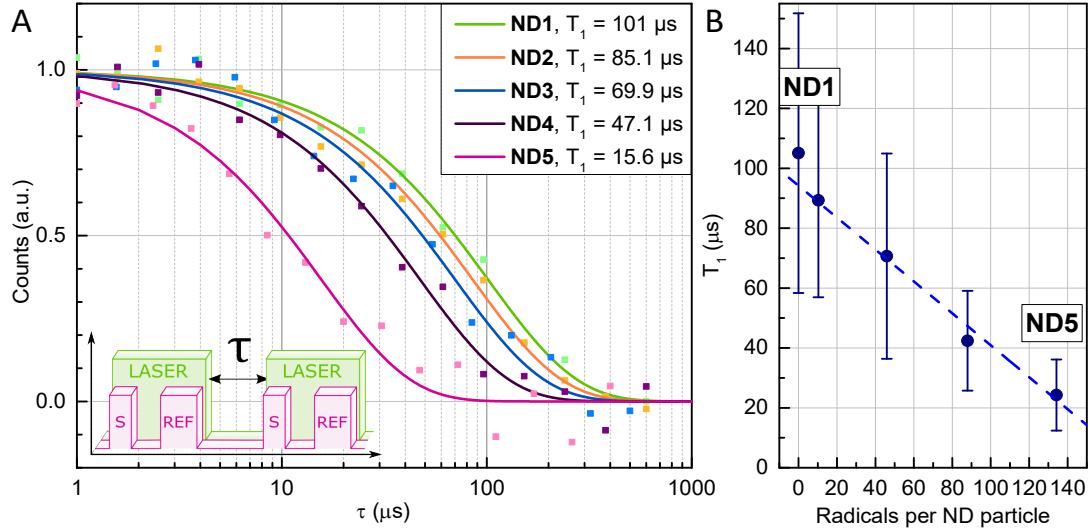

**Figure 1:** A) Determination of the $T_1$ relaxation time of single **ND1**-**ND5** particles. The pulse sequence is shown in the inset. Consecutive laser pulses are used for spin-initialization and readout, while the waiting time $\tau$ is increased. The spin-contrast signal is detected optically at the beginning of the laser pulses (S) and is referenced to the polarized state signal (REF) at the end of the laser pulses. B) Dependence of the average $T_1$ relaxation time of NV centers on the load of stable nitroxide radicals on the polymer interface of NDs. The dashed line shows a linear fit of the data. The error bars represent the standard error from the $T_1$ distribution within the different particles measured.

**Simulation of $T_1$ distribution.** To understand in-depth the relation between the $T_1$ spin relaxation and the number of radicals present in the polymer layer on NDs (Figure 1), we derived a theoretical model describing their interaction. The $T_1$ of a central NV electron spin in an ND is determined by its intrinsic NV relaxation and the relaxation caused by environmental noise. For the latter, the radicals in the polymer sensed by the NV generate a fluctuating magnetic field at the location of the NV center. This statistically fluctuating magnetic field can be described approximately by an amplitude variance $B_\perp^2$ and the temporal correlation time $\tau_c$ of the magnetic fluctuations, increasing the NV relaxation rate by the amount[23]

$$\frac{1}{T_1^{\text{radical}}} = 3(\gamma_e B_\perp)^2 \frac{\tau_c}{1+(\omega_{NV}\tau_c)^2}, \qquad (1)$$

where $\omega_{NV} \approx 2\pi \times 2.87$ GHz represents the NV electron spin resonance frequency and $\gamma_e$ is the electron gyromagnetic ratio. In our analysis, we also considered intrinsic relaxation processes. For example, phonons in the diamond cause relaxation rate $1/T_1^i$, which can be estimated from the values of NVs in bulk diamond. The electromagnetic noise from the surface of the ND also decreases the relaxation times of NV electron spins, resulting in smaller relaxation time for a smaller ND particle. To model this relaxation process, we included relaxation rate $1/T_1^{\text{noise}}$ associated with a layer of electron spins nearby the surface of the ND. From the total relaxation rate of each NV center,

$$\frac{1}{T_1} = \frac{1}{T_1^i} + \frac{1}{T_1^{\text{noise}}} + \frac{1}{T_1^{\text{radical}}} \qquad (2)$$

we obtained a simulated distribution of $T_1$ (see Materials and methods for details). The results of simulation including the size distribution obtained from TEM images revealed an excellent correlation between the distribution of experimentally measured $T_1$ and the corresponding calculated values (Figure 2). The simulation confirmed a broad natural distribution of $T_1$ for particles without nitroxide radicals (**ND1**), ranging mostly from around 50 to 150 µs with some extremes at 200 µs. The distribution was broadened by size and shape polydispersity of NDs, different numbers of lattice defects in individual particles,[53] and potentially by other factors such as variations in individual polymer thicknesses. In the presence of radicals, a much narrower distribution was obtained (**ND5**) with the $T_1$ time peaking at approximately 20 µs. These broad distributions are reflected in the relatively large error bars in Figure 1B.

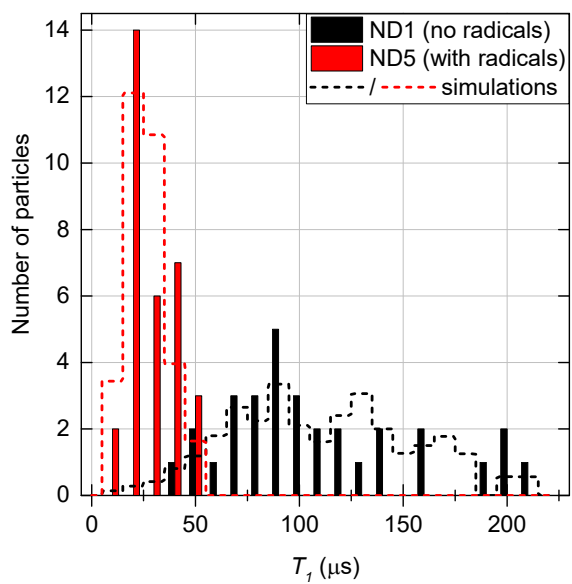

**Figure 2:** $T_1$ histograms for samples **ND1** (without nitroxide radicals) and **ND5** (134 radicals per particle on average). $T_1$ values measured on single particles are shown as bars. Histograms obtained from the corresponding theoretical simulations appear as dashed lines.

**Dynamic monitoring of a redox chemical process.** The experimental results and theoretical simulations confirm the possibility of sensitive detection of a small number of radicals in the ND vicinity. We further focused on employing these NDs as a probe for dynamical detection of a biologically relevant redox reactions (e. g. products of redox homeostasis) by monitoring the $T_1$ change during the radical reduction in the ND polymer shell. Cells maintain redox homeostasis by a series of different mechanisms[56] that depend on enzymatic and non-enzymatic antioxidants such as ascorbic acid (vitamin C), glutathione, and melatonin.[34] Ascorbic acid is a particularly important natural antioxidant occurring in plants and is an essential component of human nutrition. Ascorbate anions, the dominant form of ascorbic acid under physiological conditions, function as a redox buffer to maintain homeostasis of eukaryotic cells.[57] Ascorbate is also involved in a range of other processes, including tissue regeneration, regulation of enzyme activities,[58] and regulation of epigenomic processes.[59]

Developing biocompatible and bioinert nanoscale tools for sensitive detection of ascorbate could aid understanding of the role of ascorbate within the cells.[60–62] Because ascorbate itself does not have a free spin electron, selective chemical reactions are needed for its detection. Paramagnetic nitroxide radicals, especially TEMPO-based structures, react with ascorbate in a well-defined manner, providing diamagnetic cyclic hydroxylamines (Scheme 1C).[63] Reactions of ascorbate with nitroxides have been used to construct analytical tools for fluorescent[64,65] and EPR detection[66–68] of ascorbate. Although fluorescent probes enable sensitive and localized detection of ascorbate both in cells[65,69] and in vivo,[64,70–73] they have certain disadvantages, such as photobleaching and a pH-dependent response. EPR-based methods are very sensitive, but their resolution is fundamentally limited[74] to measurements of relatively large volumes. In contrast, our technique provides detection with high spatiotemporal resolution (individual NDs, minute time-scale) without photobleaching (NV centers have nearly unlimited photostability[75]).

To mimic the intracellular sensing events of nitroxides magnetically coupled with NV centers in NDs, we investigated the redox reaction occurring between nitroxide radicals present in the polymer layer and free ascorbate (Scheme 1C). In this proof-of-concept experiment at the single-particle level, we measured by our nanosensor temporal progression of $T_1$ after the addition of ascorbate to the liquid medium. We drop-casted an aqueous solution of **ND5** on a clean cover glass mounted on a confocal microscope. We selected a random **ND5** particle attached to the cover glass and optically measured $T_1$. Immediately after this initial measurement, ascorbic acid was added to the droplet, and changes in $T_1$ caused by reduction of nitroxide radicals to diamagnetic hydroxylamines (Figure 3A) were measured over time (see Materials and methods for details). By using an approximately 220 μM concentration of ascorbate, we could monitor the gradual completion of the redox reaction at the ND surface within a few minutes. The $T_1$ traces were averaged for 1-2 minutes, which was a limit to get sufficient signal-to-noise. The reaction proceeded with first-order kinetics and a rate constant $k$ of roughly 0.01 s$^{-1}$ (Figure 3B). Due to the fast kinetics of the reaction, the $T_1$ time changed during the acquisition time increasing the fitting error. However, the time-dependent trend can be observed clearly. Data presented in Figures 3A and 3B are representative results from measurements on one single ND. Similar results were obtained when the experiments were repeated for other **ND5** particles. These measurements demonstrate the possibility of monitoring redox reactions in real-time on a minute scale by means of stable radical probes coupled with NV centers.

To confirm that the observed changes in $T_1$ were caused by nitroxide reduction, we independently measured bulk solutions of **ND5** using EPR before and 1 h after the reaction with ascorbate. While ND spin defects remained unchanged, we observed a complete disappearance of the characteristic nitroxide triplet (Figure 3C) after ascorbic acid addition. The ND spin defects, including unpaired electrons mostly located between 0.4 and 1 nm from the surface,[76] cannot be reduced by ascorbate, because they are not sterically accessible to bulky reduction agents such as ascorbate.

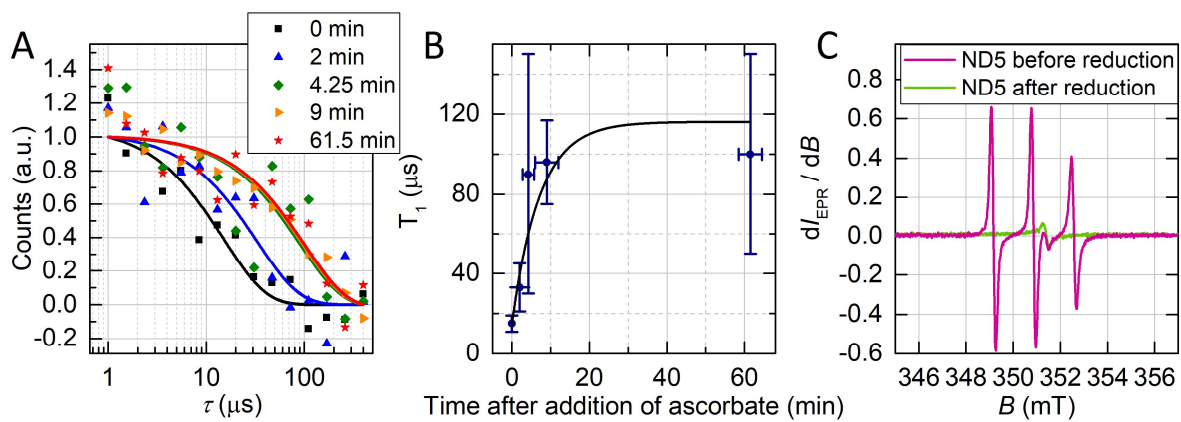

**Figure 3:** A, B) Response of NV $T_1$ relaxation time to reduction of polymer-bound nitroxide radicals with ascorbic acid to diamagnetic hydroxylamine. A) Determination of $T_1$ relaxation time of a single **ND5** particle at different time points after the addition of ascorbic acid. The counts were normalized. B) Gradual development of $T_1$ after the addition of ascorbic acid. The error bars in the direction of x-axes represent the overall $T_1$ measurement time. The error bars in y-axes represent the $T_1$ fitting error. C) EPR spectra of **ND5** with attached nitroxide radicals before and after the addition of ascorbic acid. While the nitroxide radical triplet completely disappears, the stable lattice spin defects in ND occurring at approximatelly 351.4 mT remain unchanged.

**CONCLUSION**

Improving the understanding of the role of redox processes in maintaining cell homeostasis is of key importance and requires the development of sensitive tools that provide spatial and temporal information. To address this need, we designed and constructed a first-generation relaxometric nanosensor based on colloidally stable ND particles bearing a polymer layer with nitroxide radicals. We used confocal $T_1$ electron spin relaxometry based on magnetic interaction between NV centers and nitroxides in single individual NDs. Our results revealed that $T_1$ NV relaxometry is capable of detecting the presence of stable nitroxide radicals covalently bound to a single ND particle. We reached a sensitivity of ~10 single radical spins per ND particle, which corresponds to a localized readout of approximately $10^{-23}$ mol of radicals. Additionally, to mimic a cellular environment and demonstrate a crucial step of dynamic readout of a chemical reaction, we applied our method to quantitatively assess the reduction of nitroxide radicals on NDs by ascorbate in aqueous solution at ambient temperature. We monitored the reaction with a minute resolution, thus validating the potential to use colloidally stabilized ND-radical conjugates as a sensitive nanosensor for localized detection of antioxidants. To further reduce measurement time and therefore increase the temporal resolution of the dynamic measurements, additional ND surface chemistry modifications could be made to increase the NV coherence time,[77] a recently demonstrated[78] enhanced readout method utilizing spin-to-charge conversion could be employed, and/or NDs with an ultrahigh concentration of NV centers could be used. Further improvement in sensitivity can be achieved using double electron-electron resonance sensing which is capable to detect a single electron spin.[13]

## MATERIALS AND METHODS

**Preparation of fluorescent NDs.** NDs (MSY 0–0.05, Microdiamant, Switzerland) were treated according to a previously described procedure.[79] Briefly, NDs were oxidized by air in a furnace at 510 °C for 4 h; treated with a mixture of HF and HNO$_3$ (2:1) at 160 °C for 2 days; and washed with water, 1 M NaOH, water, 1 M HCl and 5x water until all remaining acid was completely washed out. Purified ND powder was irradiated in an external target holder for 21 h with a 16.6 MeV electron beam (1.25 × 10$^{19}$ particles cm$^{-2}$) extracted from an MT-25 microtron.[80] The irradiated material was annealed at 900 °C for 1 h [81] and subsequently oxidized for 4 h at 510 °C. The resulting powder was again treated with a mixture of acids and washed with NaOH, water, HCl and 5x water, providing a grey powder upon lyophilization of oxidized fluorescent NDs (**ND0s**). We did not estimate the exact concentration of NV centers in the **ND0**, however, its brightness was similar to our previous ND samples.[53,54] We expect that **ND0** contained roughly one to few NV centers per particle.

**Purification of NDs after surface modifications.** The following general procedure was used for ND washing: particles were spun down at 15,000-35,000 g for 25-40 min in 1.5-ml Eppendorf tubes. After reactions, NDs were spun down at 15,000-20,000 g for 20-30 min, and the rotational force and time were increased with decreasing concentration of reagents. Centrifugation was followed by checking with a green laser pointer for remaining particles in solution. If no laser trace was observed, the supernatant was removed, 1.4 ml of solvent (water, DMSO, or MeOH) was added and particles were resuspended by sonication in the cup horn.

**Polymer coating terminated with alkyne functional group (Scheme 2).** A 20-mg portion of **ND0s** was weighed into a 2-ml Eppendorf microcentrifuge tube. Glycidol (1.44 ml), freshly distilled under reduced pressure (1.5 mbar, 50 °C), was pipetted into the tube with **ND0s**. The tube was sealed and sonicated for 1 h in pulse mode (2/2 s on/off) inside of a water-cooled cup horn. The colloid solution was transferred into a pressure tube stirred at 200 rpm on a magnetic stirrer with an oil bath heated at 130 °C. After 3 h, glycidyl propargyl ether (990 µl) was added to the reaction mixture. The reaction was heated for 16 h under the same conditions. The reaction mixture was cooled down to approximately 60 °C and diluted with MeOH. The tube content was split into 20 microcentrifuge tubes (1.5 ml). The **ND0s** were washed 3x in MeOH and 3x in water. The stable colloid solution of polymer-modified **ND0-PG** was obtained and stored in the dark at 4 °C at a concentration of 10 mg/ml without apparent loss of colloidal stability for approximately one year.

**Attachment of nitroxide radical to ND0-PG (Scheme 2).** Polymer-coated **ND0-PG** were modified with **3** (see Supporting Information for synthesis) using Cu(I)-catalyzed azide−alkyne cycloaddition (click chemistry).[82] The degree of modification was controlled by different concentrations of **3** in the reaction mixture (5/50/150/500/1000 mM), while the remaining conditions were kept constant. The reaction was performed in MilliQ water with 40% DMSO. Compound **3** was dissolved in DMSO; the remaining stock solutions were prepared in water. The reactants were mixed in a 1.5-ml Eppendorf tube in the following order to a final volume of 500 µl with the following final concentrations: 0.5 mg **ND0-PG** with alkyne-terminated polymer (1 mg/ml), DMSO, **3** (5/50/150/500/1000 µM), premixed aqueous solutions of CuSO$_4$·5H$_2$O and BTTES (0.1 mM and 0.5 mM), water, aminoguanidine (1 mM), 10x PBS (1x PBS, pH 7.4), and freshly prepared solution of sodium ascorbate (1 mM). The reaction mixture was mixed and left without stirring in a sealed tube for 20 min. The reaction was stopped by addition of 120 µl of 20 mM of EDTA, pH 7.4, into 500 µl reaction volume. The NDs were washed in a 1.5-ml Eppendorf tube with 40% DMSO at 15,000 g for 15 min and 20,000 g for 20 min. Then, 24 µl of 20 mM EDTA was added into the tube and after 10 min, NDs were washed two times with 40% DMSO and four times with 0.1 mM phosphate buffer, pH 7.4. Modified NDs were stored in 0.1 mM phosphate buffer, pH 7.4, at a concentration of 10 mg/ml. The covalently bound diamagnetic **3** was converted to nitroxide radical by open-air sonication of modified NDs for 20 min (pulse 2/2 s on/off) in a water-cooled cup horn (Cole-Parmer), providing nanodiamonds **ND1-5**.

**Nanoparticle characterization.** The hydrodynamic diameters and colloidal stability of NDs were determined by **DLS** (Zetasizer Nano ZSP, Malvern Instruments) and processed with Zetasizer Software 7.13. Prior to dilution, the samples were centrifuged for 1 min at 1000 g. The samples were then diluted to a concentration of 0.05 mg/ml with MilliQ water or PBS (1X, 7.4 pH). Hydrodynamic peak size was calculated from the intensity autocorrelation function using the general-purpose algorithm in the Zetasizer software. DLS data were collected at a backscatter angle of 173° (NIBS system) at 25 °C; viscosity was set at 0.9236 and 0.8872 cP for PBS and water (Table 1 and Figure S3 in Supporting Information). Each sample was measured three times with an automatic duration; the reported size represents an average value of Z-average size. Particle size distribution is displayed by intensity. The particles were stored at 4 °C in the dark for 1 year without apparent loss of colloidal stability.

Particle concentration was determined using **NTA** with a NanoSight NS300 (Malvern Instruments) equipped with a low-volume cell and 405 nm laser. For data analysis, built-in Malvern software (NTA 3.4 - Sample Assistant Build 3.4.003-SA) was used. Before dilution, samples were sonicated in the cup horn and centrifuged for 1 min at 1000 g. ND samples (approximately 10 mg/ml) were diluted 100,000 times in three steps with MilliQ water and injected using an integrated autosampler. Dilution was optimized to obtain an average of 35-90 particles per frame with 1500-5500 valid tracks per 60 s measurement at a temperature of 25 ± 0.1 °C. Each sample was diluted in duplicate and each diluted sample was measured for 10 x 60 s with syringe pump speed 100. Captured data were analyzed using a finite track length adjustment (FTLA) particle size distribution algorithm. The camera level was set to auto with the following parameters of level/shutter/gain: 11/890/146, 12/1200/146, 13/1232/219 or 14/1259/366. The detection threshold was set to level 3, maximum jump distance was between 14.20-17.50; the blur, maximum jump mode, and minimum expected particle size (MEPS) were set to "auto."

**TGA** was performed at the Laboratory of Thermogravimetric Analysis, UCT Prague, using a Stanton-Redcroft TG750 thermobalance. The sample (1-2 mg) was placed on a platinum pan as a solution (10 mg/ml), dried by heating overnight at 80 °C, and heated from room temperature to 900 °C at a rate of 10 °C/min under air purge (20 ml/min).

**Oxygen sensitive operations** with NDs were performed inside a Sekuroka glove bag. The oxygen concentration in the protective argon atmosphere in the bag during sample handling was under 300 ppm according to continual measurement with an $O_2$ Gas sensor ($O_2$-BTA, Vernier) connected to a LabQuest2 monitoring unit (Vernier). Before use, all solvents were deoxygenated by three freeze-thaw cycles. Oxygen sensitive sample handling outside the glove bag (centrifugation and sonication) was performed in Eppendorf tubes with the opening covered by parafilm to reduce possible sample oxidation.

**Infrared spectra** were measured on a Fourier Transformer Infrared Spectrophotometer (FTIR) Nicolet 6700FT-IR spectrometer (ThermoScientific, USA) with a standard MIR source, KBr beamsplitter, and MCT/A detector. The spectrometer was purged with dry nitrogen. Spectra were acquired in attenuated total reflection mode (ATR) using ATR-MIRacl$^{TM}$ – a triple-reflection ZnSe horizontal ATR prism (Pike Technologies, USA) in the spectral region 3800 – 600 cm$^{-1}$, with spectral resolution of 4 cm$^{-1}$ and 512 scans. A Happ-Genzel apodization function was used. Data processing was performed using OMNIC software.

**Transmission electron microscopy (TEM) and image analysis.** Bright-field TEM experiments were performed with a JEOL JEM-1011 electron microscope operated at 80 kV equipped with a Tengra bottom-mounted camera. Particles for TEM samples were placed on copper grids (SPI) with a home-made Parlodion membrane and carbon coating. TEM sample preparation: For naked **ND0** particles, the grid was pre-treated with a droplet of poly(ethyleneimine) solution ($M_W$ = 2.5 kDa, 0.1 mg ml$^{-1}$). The droplet was removed with a piece of tissue after 10 min incubation. Then, the grid was placed on a droplet of deionized water for 1 min and dried once again. For **ND0-PG** polymer-coated samples, the grid was used without any pre-treatment. The grid was placed on a droplet of an aqueous solution of **ND0** or **ND0-PG** (0.1 mg ml$^{-1}$) and the solution was removed with a piece of tissue after 3 min incubation. Phosphotungstic acid solution (PTA, 2%, pH 7) was used for negative staining. Samples for staining were placed without drying twice on 50 μl droplets of PTA solution for 5 min. All samples were then washed two times on a 50 μl droplet of water for 1 min and dried with a piece of tissue. The polyglycerol shell thickness was analyzed with ImageJ software from 50 particles and estimated to be 5.7 ± 1.6 nm. Each thickness was measured at one or two positions, depending on the TEM micrograph quality.

For **cryo-TEM**, samples were prepared as previously described.[83] Briefly, 3 μl of sample solution (2 mg/ml) were deposited on a QUANTIFOIL® C-flat CF-4/2-2C-50 grid, previously treated with glow discharge (2x10$^{-1}$ Torr, 10 mA, 30 s). The excess solution was removed by blotting with Whatman No 1 filter paper, and the grid was plunged into liquid ethane held at -183 °C using an EMS-002 semiautomatic plunger (Electron Microscopy Science, Hatfield, PA, USA). The samples were observed under an FEI G20 Sphera electron microscope operated at 120 kV. Images were recorded using low dose mode (less than 20 e/Å$^2$) on a Gatan US2000 slow-scan CCD camera at 29,000x direct magnification.

For image analysis – the polyglycerol layer detection – the images were first band-pass filtered (1-400 px window) to remove gradients and then Gaussian blurred with σ level adjusted to 6 (ImageJ v. 1.48 software[84]). The contours of the polyglycerol layer were then automatically detected using NIS software (NIS Elements 5.02, Nikon). The thickness of the layer (17.0 ± 8.8 nm) was determined from 30 independent measurements.

**EPR spectroscopy.** The EPR spectra of aqueous solutions of **ND1-5** functionalized with nitroxide radicals, as well as of unmodified **ND0** powder, were recorded on a Bruker EMX$^{plus}$ 10/12 CW (continuous wave) EPR spectrometer equipped with a Premium-X-band microwave bridge (both Bruker). Reduction of nitroxide radicals on the ND-surface (change from paramagnetic to diamagnetic state) was monitored on an ELEXSYS E580 6/1 FT (Fourier transform)/CW (Bruker) EPR spectrometer under CW-conditions. The $g_{iso}$ of the studied radicals was determined using a built-in spectrometer frequency counter and an ER-036TM NMR-Teslameter (Bruker). Hirschmann ringcap capillaries (inner diameter 78 mm, Hrischmann Laborgeräte GmbH, Germany) were used as sample tubes for quantitative EPR measurements. The capillaries were calibrated for a volume of 50 μl with an accuracy of ≤ ± 0.25% and precision of ≤ 0.5%. Finally, the capillaries were inserted into the high-sensitivity EPR cavity (ER-4119HS, Bruker), which was applied throughout the quantitative experiments. Samples **ND1-5** in aqueous solution were each measured 3- to 4-times to assess the reproducibility and calculate the mean value of the total double integral and/or the absolute number of spins (see EPR calculations). For such quantitative measurements, the following instrumental parameters were applied:

frequency = 9.867 GHz, central field = 351 mT, sweep width = 12 mT, power = (1.26-2.52) mW, modulation amplitude = 0.072 mT, resolution = 2500 points, conversion time = 5.8 ms, time constant = 2.6 ms. The EPR spectra of the solid **ND0** sample (the precursor used in next steps for polymer coating and nitroxide attachment; see section 3.1) were recorded eight times using the following parameters: frequency = 9.869 GHz, central field = 352 mT, sweep width = (28-32) mT, power = (0.63-3.17) mW, modulation amplitude = (0.064-0.120) mT, resolution = 1900 points, conversion time = 18.0 ms, time constant = 5.1 ms. No saturation effects were observed for either aqueous or powder samples.

**EPR calculations.** Spectra were acquired and preprocessed with Xenon 1.1b.159 software (Bruker, Germany). The preprocessing included baseline correction and double integration (DI) of the EPR spectrum. Additional analysis, within the above-described software, covered the calculation of the absolute amount of spins in samples. With the known tube geometry and including all the quantitative relevant instrumental parameters, the total amount of spins/cm$^3$ could be calculated based on the double integral of the EPR spectrum.[85] The DI requires a high signal-to-noise ratio of the experimental spectrum to achieve a well-defined spectrum integration curve, and this requirement was fulfilled for all recorded EPR spectra.

Afterwards, deconvolution/decomposition of spectra of aqueous samples into a "sub-surface" radical centered at $g = 2.00254 \pm 0.00006$ (see also[76]) and nitroxide radical (centered at $g = 2.00555 \pm 0.00004$) components was done using MATLAB toolbox EasySpin v. 5.1.[86] The analysis involved fitting/simulation of the experimental EPR spectra by varying parameters such as $g$-factor, $^{14}$N hyperfine coupling constants (HFCCs, $A$), rotational correlation time, and spectral linewidth as well as spectral weights (double integral ratios) of the "sub-surface" and nitroxide radical components. For this purpose, the "garlic" fitting function was used, because the EPR spectra of nitroxides fall into the region of fast-motion regime.[86] The observed lack of a P1 center (substitutional nitrogen, termed also $N^0$ or $N_s$ center) conventional hyperfine pattern[87] and its shift and overlap with milling-induced defects (carbon dangling bonds carrying unpaired electron spin) is typical for NDs smaller than ~80 nm [88] and results in a single spectral signature of "sub-surface" radicals.[49] Although the "sub-surface" signal at $g = 2.00254 \pm 0.00006$ comes from the solid-state paramagnetic centers within ND particles, their one-line signal could be thus accurately simulated like a radical component without any hyperfine splitting/coupling, and its line-form did not change within the sample series (Figure S6 in Supporting Information).

**Optical characterization of the particles.** A custom-built confocal microscope was used to probe the NV spin dynamics. To excite the NV centers, a solid-state laser with a wavelength of 532 nm was pulsed using an acousto-optic modulator (Crystal Optics, 3200). The laser beam (24 µW) was focused onto the sample using an oil-immersion objective (Olympus UPlanSApo 60x oil, N.A. = 1.35). Collected fluorescence was filtered by a long pass 650 nm detection filter and then detected by an avalanche photodiode with single-photon resolution (Excelitas Technologies).

The laser pulse sequence for the $T_1$ time measurements was applied using an Arbitrary Waveform Generator (Tektronix AWG70001A). The $T_1$ measurement pulse sequence consisted of repetitive 10-µs laser pulses. At the end of each pulse, the NV spin was polarized into its $m_s = 0$ state. After a variable waiting time ($\tau$), a consequent laser pulse was used to read out the NV spin state. The readout window of the signal (S) was roughly 1 µs long and was at the beginning of each laser pulse. The reference counts (REF) were collected for roughly 3 µs at the end of the pulse (see the inset of Figure 1A). The measured fluorescence data were plotted with respect to $\tau$ duration and fitted with a mono-exponential function to extract the $T_1$ longitudinal relaxation time. To increase the signal-to-noise ratio of the measurements, the sequences were repeated multiple times, amounting to a total acquisition time of 3-8 minutes for a single site, depending on the fluorescence intensity.

The glass slides were cleaned with acetone, ethanol and water, and the cover glasses were plasma cleaned to minimize background fluorescence. For $T_1$ time statistics, the samples were prepared by placing a 4 µl drop of 0.05 mg/ml ND solution in ultra-pure water between the cover glass and the glass slide. Cover glass edges were sealed with tape to avoid water evaporation. Measurements were performed in aqueous solutions immediately after sample preparation. The measurements on **ND1** and **ND5** depicted in Figure 1 are averaged from 31 and 32 particles, respectively, and the averaged $T_1$ fitting errors are 17.9 µs and 16.2 µs, respectively.

**Determination of particle concentration.** The precise weight concentration of the particles (and the corresponding amount of polymers) used for EPR measurements was estimated by a combination of TGA and NTA analyses. First, the precise weight concentration and the polymer/ND weight ratio was estimated using TGA. A known volume of **ND0-PG** stock solution was concentrated to approximately 20 mg/ml (of known weight). A 50-µl aliquot of this solution was placed in a balanced thermogravimetric cup, slowly dried overnight at 80 °C, and subjected to TGA (RT to 800 °C; 10 °C/min; Figure S3).

For NTA analysis, stock solutions of **ND1-5** and **ND0-PG** (approximately 10 mg/ml) were diluted 100,000 times and measured. The obtained number-weighted concentration of NDs (typically in the range: 3.9-4.9·$10^{13}$ particles/ml) was then recalculated using the volume-weighted distribution of ND based on TEM image analysis,[89] providing the weight concentration of the ND component of **ND0-PG** (mg/ml). The final weight concentration of the particles was determined from the known weight ratio of polymer and ND component in **ND0-PG** estimated by TGA.

**Determination of radical concentrations in NDs.** The concentrations of radicals in **ND1-5** ($g^{-1}$) were obtained by recalculating the radical volume concentration determined by EPR ($cm^{-3}$) to weight concentration in the individual samples estimated by a combination of NTA and TGA. Values are listed in Table S1 in Supporting Information. As a control for our results from an aqueous environment, in which discrimination of the P1 signal can potentially occur,[90] EPR spectra of a solid **ND0** sample were recorded. The concentration of paramagnetic centers (centered at $g = 2.00258 \pm 0.00007$) was estimated as $(2.53 \pm 0.08) \cdot 10^{18}\,g^{-1}$. This value correlates well with the concentration range determined in water (Table S1 in Supporting Information). The EPR simulations as well as comparison of the solid-state and aqueous spectra are shown in Figure S6 in Supporting Information.

**The nitroxide radical loads in** ND1-5 **(number of radicals per particle).** As estimated by cryo-TEM, the average thickness of the polymer layer on **ND0-PG** ($17.0 \pm 8.8$ nm) was independent of the particle sizes (Figure S4). In our geometrical model, this thickness was formally added to each individual particle used for calculation of NTA/TEM distribution. The total volume of polymer present in the sample was obtained as a sum of volume contributions from all individual particles. The radical concentration obtained from EPR was recalculated to this volume and used for other considerations (Table 1 and Figure S7 in Supporting Information).

**Reaction of** ND5 **with ascorbic acid and related $T_1$ and EPR measurements.** For measurements of $T_1$ time recovery, the sample was prepared by placing 4 µl of ND solution directly on a cover glass without sealing. After the initial $T_1$ measurement, 3 µl of the 0.1 mg/ml solution of ascorbic acid was added to the ND solution droplet on the cover glass without moving the sample. $T_1$ measurements with acquisition time of 1-2 minutes were repeated on the same ND particle for different waiting times to inspect the $T_1$ time-dependency and continued until the $T_1$ stabilized (Figure 3A, B).

For control EPR measurement, 75 µl **ND5** solution (10 mg/ml) was transferred into an Ar-filled glove bag and thoroughly degassed. 25 µl ascorbic acid (500 mM) was added and quickly mixed. An EPR measurement was started within 5 min after mixing the sample with ascorbic acid (Figure 3C and Figure S6B in Supporting Information). The following instrumental EPR settings were used: frequency = 9.851 GHz, central field = 351 mT, sweep width = 12 mT, power = 2.993 mW, modulation amplitude = 0.075 mT, resolution = 1024 points, conversion time = 14 ms, time constant = 2.6 ms.

**Simulations.** Using the results of Tetienne et al.,[91] we considered the variance of radical-induced magnetic field $B_\perp^2 = \sum_j B_{\perp,j}^2$ to be a sum of the terms

$$B_{\perp,j}^2 = \left(\frac{\mu_0 \gamma_e}{4\pi}\right)^2 \frac{S(S+1)}{3} \left(\frac{5 - 3\left(\hat{r}_{c,j} \cdot \hat{z}\right)^2}{r_{c,j}^6}\right)$$

from all fluctuating radicals. Here, spin $S = \frac{1}{2}$, and $r_{c,j} \hat{r}_{c,j}$ (with $|\hat{r}_{c,j}| = 1$) is the position of the radical relative to the NV center. We chose $\hat{z}$ as the unit axis along the NV symmetry axis, and the summation in $B_\perp^2$ became an integral by assuming a volume density $\rho$ of the radicals. Following the procedure of Tetienne et al.,[91] we obtained

$$\tau_c = \left(\frac{\mu_0 \gamma_e^2}{4\pi} \frac{S(S+1)}{3} \frac{(8\pi\rho)^{1/2}}{r_{min}^{3/2}}\right)^{-1},$$

for the correlation time of the temporal fluctuations of the magnetic field generated by the radical spins at the position of the NV center, where $r_{min}$ is the minimum allowed distance between the radicals.

Similar to the case of a fluctuating magnetic field from radicals, the spin noise from the surface of the ND was given by

$$\frac{1}{T_1^{noise}} = 3(\gamma_e B_\perp^{noise})^2 \frac{\tau_c^{noise}}{1 + (\omega_{NV} \tau_c^{noise})^2},$$

using the variance $(B_\perp^{noise})^2$ and the correlation time $\tau_c^{noise}$ of the magnetic field from the surface electron spins. In this model we considered the surface unpaired electron spins as the main source of noise, neglecting the presence of P1 centers.

To simplify our Monte Carlo $T_1$ simulation, we assumed all the NDs to be spherical, with size distribution given by the experimentally measured values. Because NV centers are not stable when they are very close to the surface, we introduced a depletion zone of 1 nm close to the surface.[83,92] NV centers that contribute to the detected signal can be anywhere in the diamond except the depletion zone. We assumed that the noisy surface electron spins are distributed homogeneously on the surface of the ND with a 1-nm-thick shell. For radical sensing, we added a polymer layer of 10 nm surrounding the ND closely with a homogeneous radical density. Because the radicals couple randomly to the NV *via* dipolar interactions, the relaxation is dominant for the radicals close to the NV center (see the expression of $B_{\perp,j}^2$). This is consistent with our observation that the simulation had a similar result when a thicker homogeneous polymer layer was used (not shown).

In the simulation, we used $r_{min} = 0.15$ nm, following Tetienne et al.,[91] and $T_1^i = 1$ ms, which is close to the typical NV lifetime in bulk diamond at room temperature. We found that the trend of $T_1$ distribution is not sensitive to these values. For simulations without the polymer layer, the density of the noisy electrons on the ND surface was chosen to be 0.51 nm$^{-3}$ to make the simulated $T_1$ distribution fit the experimental data. The same density of the noisy electrons was also used for simulations including a polymer layer. A radical density of 1.1 nm$^{-3}$ was used for the polymer layer to obtain the distribution shown in Figure 2 in the main text. These densities are similar to those obtained by Tetienne et al.[91] Because experimental settings include noise sources such as surface-modified phonons that increase the relaxation rate of NV centers in a form similar to the expression of $1/T_1^{\text{noise}}$,[93] we expect the electron densities used for our simulation to be higher than experimentally determined electron densities.

## ACKNOWLEDGEMENTS

The authors are grateful to Priyadharshini Balasubramanian for kind providing the optical setup for the relaxometry measurements, to David Chvatil for electron irradiation of the NDs, and to Lucie Bednarova for measurement of IR spectra. This work was supported by the Czech Science Foundation project Nr. 18–17071 S (to P.C.), MSM project Nr. 8C18004 (NanoSpin) (to J.B., H.R., P.C., J.S.), Charles University, project GA UK No. 860217 (to J.B., P.C.), European Regional Development Fund; OP RDE; Projects: Chem-BioDrug (No. CZ.02.1.01/0.0/0.0/16_019/0000729) (to P.C., H.R.) and CARAT (No. CZ.02.1.01/0.0/0.0/16_026/0008382) (to J.B., H.R., P.C., J.S.), Flemish Scientific Research Project FWO (No. G0E7417N; to M.N., M.G), QuantERA project Q-Magine (No. R-8843; to M.N, F.J.), EuroNanoMed II project NanoBit (No. 258074; to M.N.), ERC Synergy grant BioQ (Grant No. 319130), HyperQ (Grant No. 856432), EU H2020 Quantum Technology Flagship project ASTERIQS (Grant No. 820394), EUH2020 Project Hyperdiamond (Grant No. 667192), DFG via a Reinhardt Kosseleck project, BMBF via NanoSpin and DiaPol (to M.B.P., F.J. and Z.Y.W.). F.J. acknowledge support of VW Stiftung, DFG (via CRC 1279 and EXC 2154 POLiS) and BW Stiftung. Electron irradiation was supported through Czech Academy of Sciences project No. RVO61389005.

# Supporting Information

**Chemical synthesis**
**General**
Chemicals and solvents used for organic synthesis were supplied by Sigma-Aldrich, Penta, Lachema, Fluka, and Acros Organics and used as received unless stated otherwise. (2,2,6,6-Tetramethyl-4-piperidinyl)acetic acid hydrochloride was obtained from Chembridge Corporation. Inert operations were performed under a protective atmosphere with nitrogen or argon gas (99.999%), using vacuum line and Schlenk techniques. Water was purified using a MilliQ system.

Chromatography purifications were performed by HPLC using a 515 HPLC pump with 2996 PDA (Waters) and TOY18DAD800 compact preparative system (ECOM spol.). The following reverse phase columns were used: ReproSil Gold 120 C18, 10 μm, 250 x 20 mm (length x inner diameter) with guard 30 x 20 mm (column volume (CV) = 60 ml, flowrate 20 ml/min) and ReproSil Gold 120 C18, 10 μm, 250 x 40 mm with guard 50 x 40mm (CV = 250 ml, flowrate 80 ml/min). Compounds were detected by UV-absorption at 210 nm, and their identity was confirmed by MS. The purification started with 95% MilliQ water with 0.1% TFA and 5% HPLC methanol; the sample was injected onto the column in DMSO. The purification gradient (in % of MeOH) is described by CV and is independent of systems or columns. Methods: isocratic 2CV at 5% MeOH; step 0.5CV to 15%; then gradient 20CV to 30% (Method 30%) or 50% (Method 50%); followed by step 0.5CV to 100% and regeneration wash 2CV at 100%.

The nanoparticles were dispersed using a Cole Parmer Cup Horn Ultrasonic Processor Sonicator or Elmasonic P60H bath. Centrifugations were performed with Eppendorf centrifuge 5430, Sigma 3-30KS, and Eppendorf MiniSpin.

**Abbreviations used in the text**: rt (room temperature), DMSO (dimethyl sulfoxide), MeOH (methanol), TFA (trifluoroacetic acid), ND (nanodiamond), N.A. (numerical aperture).

**Chemical analysis**
HPLC-MS analysis was performed using a Waters Acquity system with a QDa detector with an Acquity UPLC BEH C18 column, 1.7 μm (100 x 2.1 mm), flow 0.5 ml/min, using 6 min gradient from 100% aqueous phase with 0.1% formic acid to acetonitrile (1.25 min).
HR MS spectra were measured on an LTQ Orbitrap XL (Thermo Fisher Scientific) instrument.
NMR spectra were recorded at room temperature in methanol-$d_4$ solutions using a Bruker Avance III™ spectrometer equipped with a Prodigy cryo-probe operating at 401.0 MHz for $^1$H and 100.8 MHz for $^{13}$C. For structure elucidation, 1D and 2D correlation NMR experiments (HSQC, HMBC) were used. All chemical shifts (δ) are given in ppm and coupling constants in hertz [Hz]. For $^1$H and $^{13}$C NMR measurements in CD$_3$OD, spectra were referenced to the solvent peak ($\delta_H$ = 3.310, $\delta_C$ = 49.00). The following abbreviations are used to express signal multiplicities: s (singlet), d (doublet), t (triplet) and m (multiplet). Data were processed with Mestrenova (v. 14).

**Synthesis of 1 and 2**

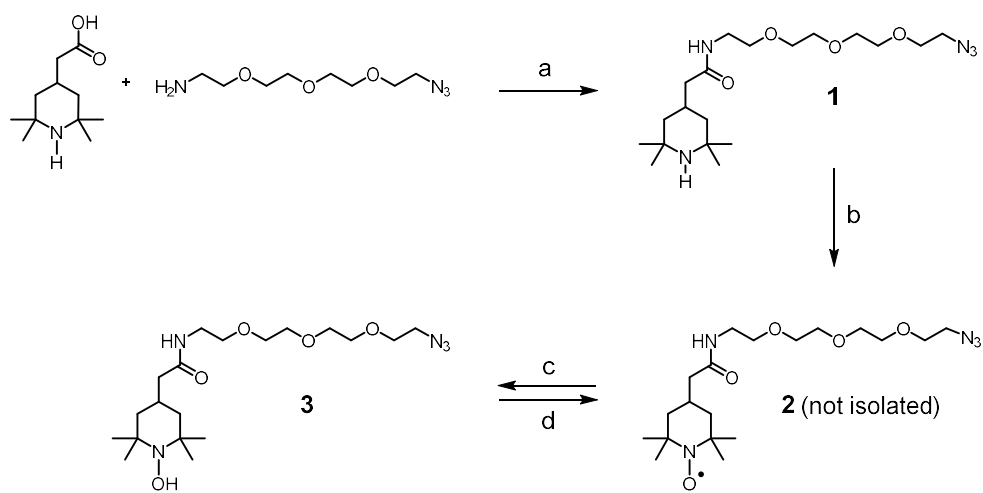

**Scheme S1:** Synthetic scheme for the preparation of TEMPO-type stable nitroxide radical **2** bearing an azide for click attachment. Nitroxide **2** can be reversibly converted to diamagnetic hydroxylamine **3**. Reaction conditions: a) HBTU coupling reagent; diisopropylethylamine (DIPEA); b) 3-chloroperoxybenzoic acid (*m*CPBA); c) sodium ascorbate; d) *m*CPBA.

**N-(2-{2-[2-(2-Azidoethoxy)ethoxy]ethoxy}ethyl)-2-(2,2,6,6-tetramethylpiperidin-4-yl)acetamide (1)**.
To a 20-ml vial equipped with a magnetic stirrer, 2-(2,2,6,6-tetramethylpiperidin-4-yl)acetic acid hydrochloride (499 mg, 2.12 mmol, 1.00 eq), 1-[2-(2-aminoethoxy)ethoxy]-2-(2-azidoethoxy)ethane (591 mg, 2.71 mmol, 1.28 eq) and dry DMSO (10.0 ml) were added, followed by addition of HBTU (924 mg, 2.44 mmol, 1.15 eq). The vial was purged with argon and dry *N,N*-diisopropylethylamine (1.5 ml, 8.47 mmol, 4.00 eq) was added *via* septum. The reaction was stirred overnight at rt. The crude reaction mixture was diluted with 3 ml of acetic acid and filtered over a 0.22-μm PTFE filter. The crude product was purified by HPLC using Method 50%. Pooled fractions were lyophilized, yielding **1** (530 mg, 62.7%) as a light-yellow oil.
**$^1$H NMR** (Methanol-$d_4$, 401 MHz) δ 3.71 – 3.58 (10H, m), 3.56 (2H, t, $J$ = 5.4 Hz), 3.42 – 3.33 (4H, m), 2.51 – 2.35 (1H, m), 2.20 (2H, d, $J$ = 7.0 Hz), 1.87 – 1.78 (2H, m), 1.47 (6H, s), 1.41 (6H, s), 1.26 (3H, t, $J$ = 13.2 Hz). **$^{13}$C NMR** (Methanol-$d_4$, 101 MHz) δ 173.77, 71.67, 71.59, 71.49, 71.21, 71.11, 70.52, 58.18, 58.12, 51.77, 49.64, 49.42, 49.21, 49.00, 48.79, 48.58, 48.36, 42.85, 42.31, 40.37, 30.69, 26.73, 25.02. **HR MS** (ESI): (+) m/z calc. for [$C_{19}H_{38}O_4N_5$]$^+$: 400.29183, found: 400.29185.

**N-(2-(2-(2-(2-Azidoethoxy)ethoxy)ethoxy)ethyl)-2-(1-hydroxy-2,2,6,6-tetramethylpiperidin-4-yl)acetamide (3) and intermediate**
**(4-{[(2-{2-[2-(2-azidoethoxy)ethoxy]ethoxy}ethyl)carbamoyl]methyl}-2,2,6,6-tetramethylpiperidin-1-yl)oxidanyl (2)**.
Compound **1** (100 mg, 250 μmol, 1.00 eq) in a 20-ml vial with a magnetic stirrer was dissolved in 4 ml dichloromethane, followed by addition of 3-chlorobenzene-1-carboperoxoic acid (*m*CPBA; 86 mg, 501 μmol, 2.00 eq). The reaction mixture was stirred overnight and the presence of **2** was confirmed by LC-MS and HR-MS.
**MS** (ESI): (+) m/z calc. for [$C_{19}H_{37}N_5O_5$]$^+$: 415.279, found: 415.279, **HR MS** (ESI): (+) m/z calc. for [$C_{19}H_{36}O_5N_5Na$]$^+$: 437.26087, found: 437.26088.
Excess sodium ascorbate (88.2 mg, 0.500 mmol, 2.00 eq) in a mixture of water and MeOH was added to the reaction mixture. It was concentrated after 1 h, dissolved in DMSO, and filtered over a 0.22-μm PTFE filter. The crude product was purified by Method 30%. Pooled fractions were lyophilized, yielding pure **1** (26.1 mg, 26.1%) and **3** (41.2 mg, 39.6%) as a light-yellow oil.
**$^1$H NMR** (Methanol-$d_4$, 401 MHz) δ 3.70 – 3.58 (10H, m), 3.55 (2H, t, *J* = 5.4 Hz), 3.42 – 3.33 (6H, m), 2.58 – 2.40 (1H, m), 2.19 (3H, d, *J* = 7.0 Hz), 2.00 – 1.91 (3H, m), 1.57 (3H, t, *J* = 13.4 Hz), 1.46 (17H, s), 1.42 (5H, s). **$^{13}$C NMR** (Methanol-$d_4$, 101 MHz) δ 173.69, 71.66, 71.59, 71.49, 71.21, 71.11, 70.50, 69.23, 51.76, 49.64, 49.42, 49.21, 49.00, 48.79, 48.58, 48.36, 43.71, 42.22, 40.36, 28.25, 25.98, 20.26. **HR MS** (ESI): (+) m/z calc. for $[C_{19}H_{38}O_5N_5]^+$: 416.28675, found: 416.28687, $[C_{19}H_{37}O_5N_5Na]^+$: 438.26869, found: 438.26877.

# NMR spectra
## Compound 1

¹H NMR spectra of **1**.

¹³C NMR spectra of **1**.

## Compound 3

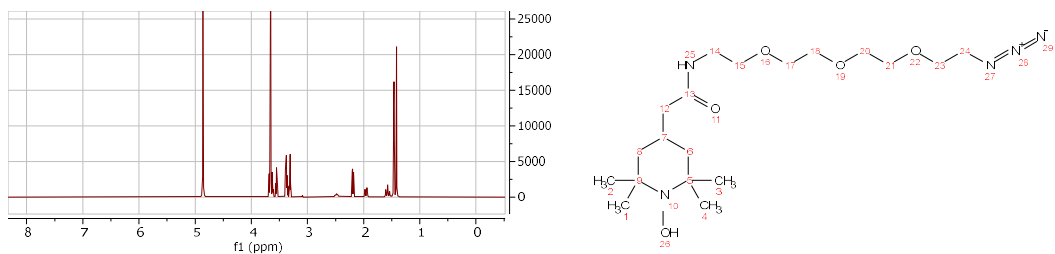

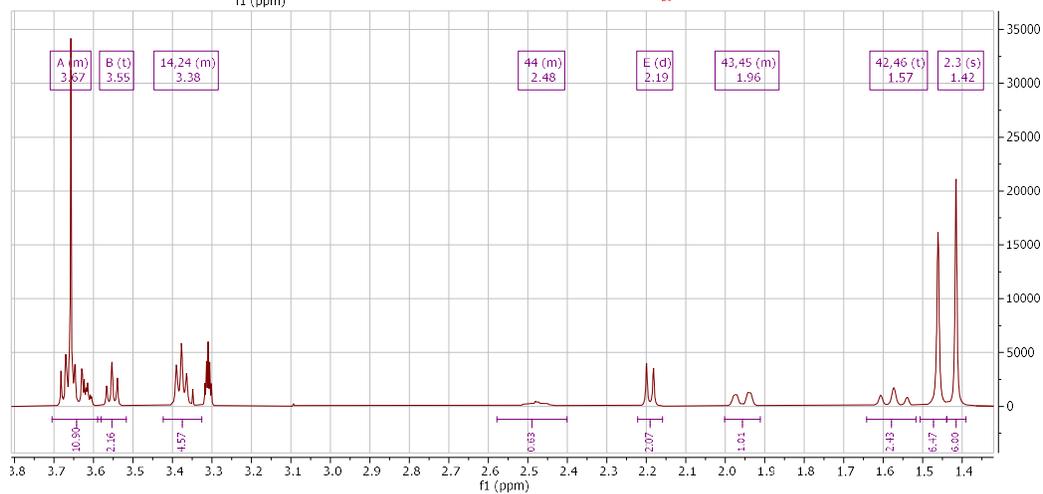

¹H NMR spectra of **3**.

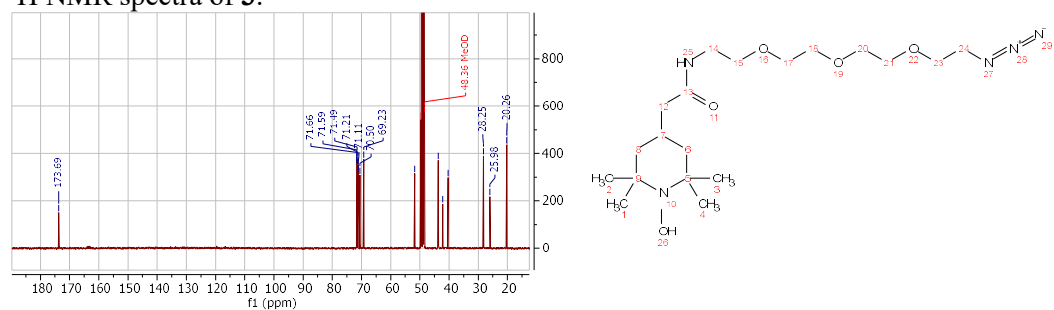

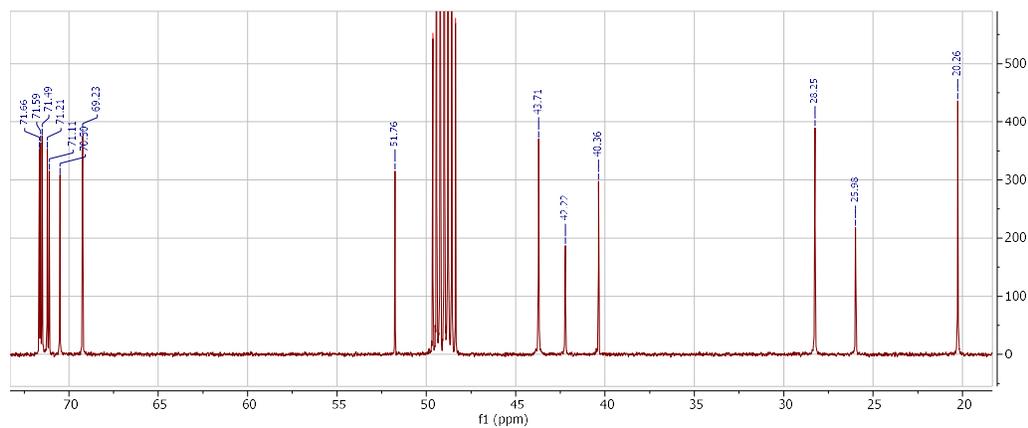

¹³C NMR spectra of **3**.

## LC-MS chromatograms and mass fragmentation
HPLC-MS chromatogram of **1-3** with different retention times (RT) on a reverse phase $C_{18}$ column using a short 8 min run.

## Compound 1

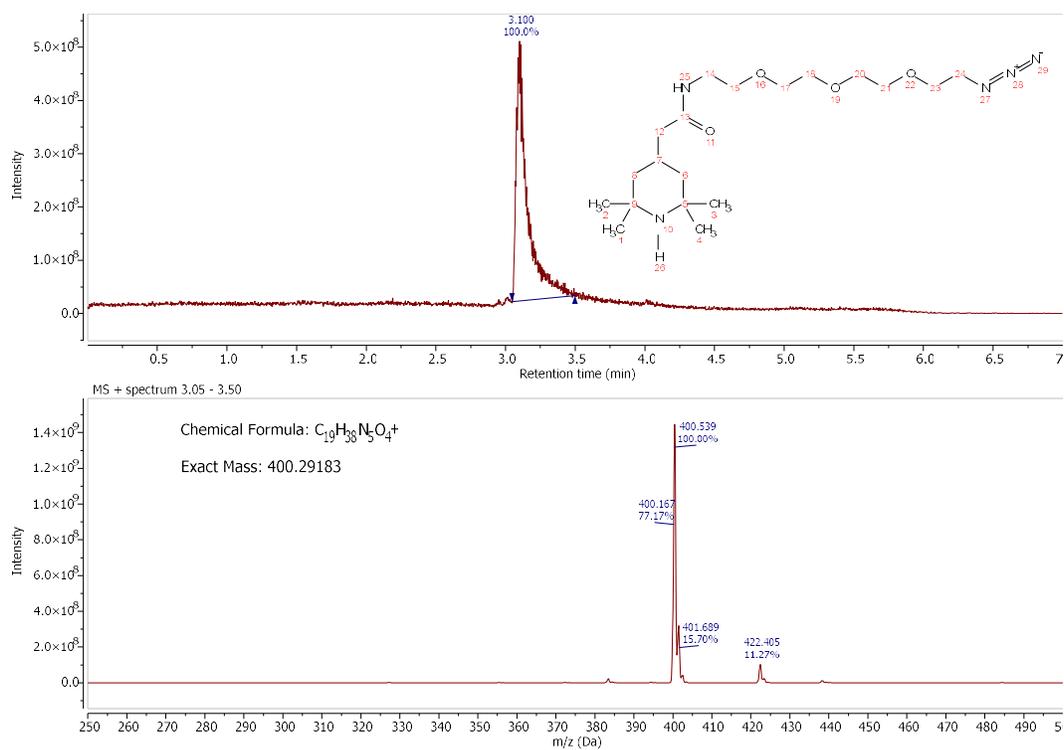

Compound **1** HPLC-MS chromatogram (RT = 3.100 min) and mass adducts $[M+H]^+$ and $[M+Na]^+$.

**Compound 2**

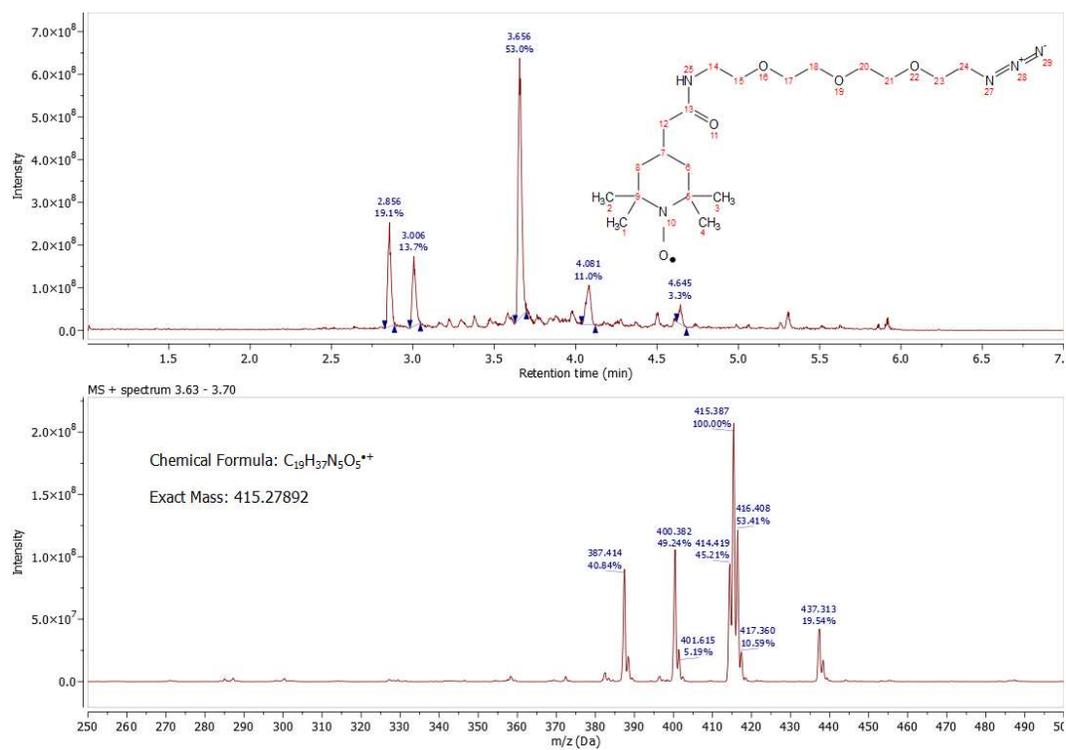

Compound **2** HPLC-MS chromatogram (RT = 3.656 min) and a more complex fragmentation corresponding to reactive radical species with adducts [M+H]⁺ and [M+Na]⁺.

**Compound 3**

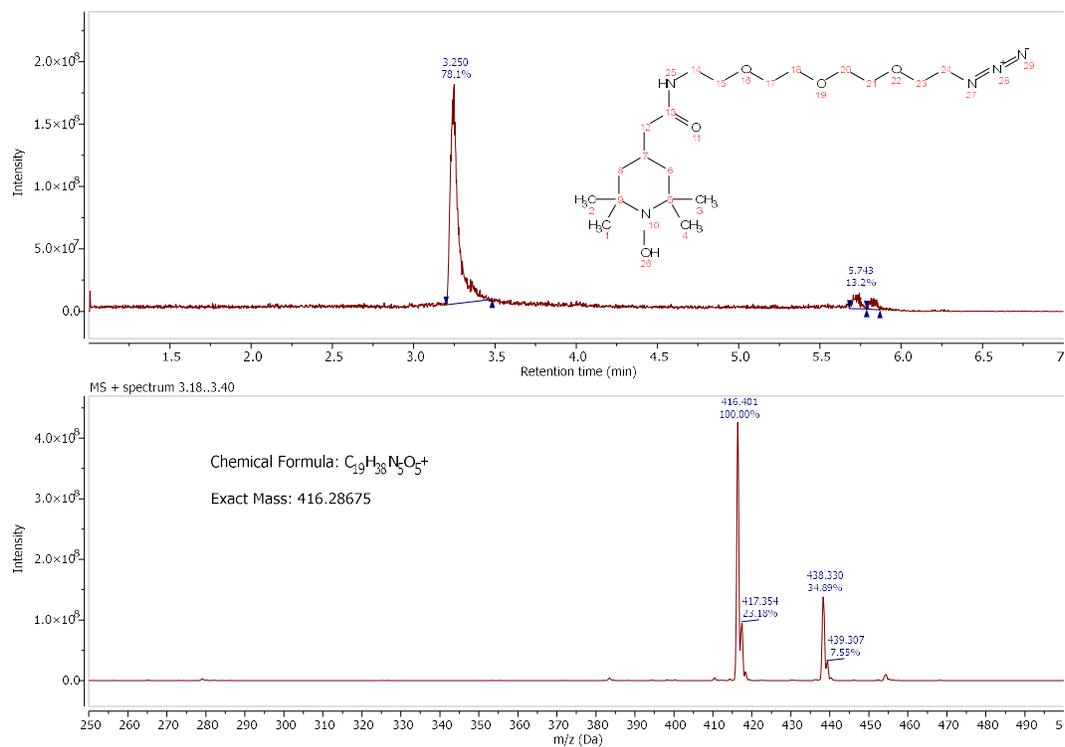

Compound **3** HPLC-MS chromatogram (RT = 3.250 min) and mass fragmentation corresponding to [M+H]$^+$ and [M+Na]$^+$ adducts.

**Additional particle characterization**

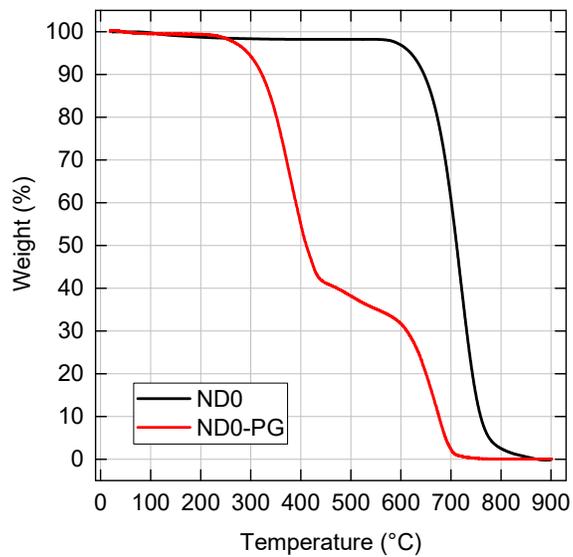

**Figure S1:** Thermogravimetric curves of **ND0** and **ND0-PG**. Decomposition of the polyglycerol layer in **ND0-PG** occurred between 200-600 °C; 64% of particle weight was assigned to the polymer. NDs decomposed in both samples between 600-800 °C.

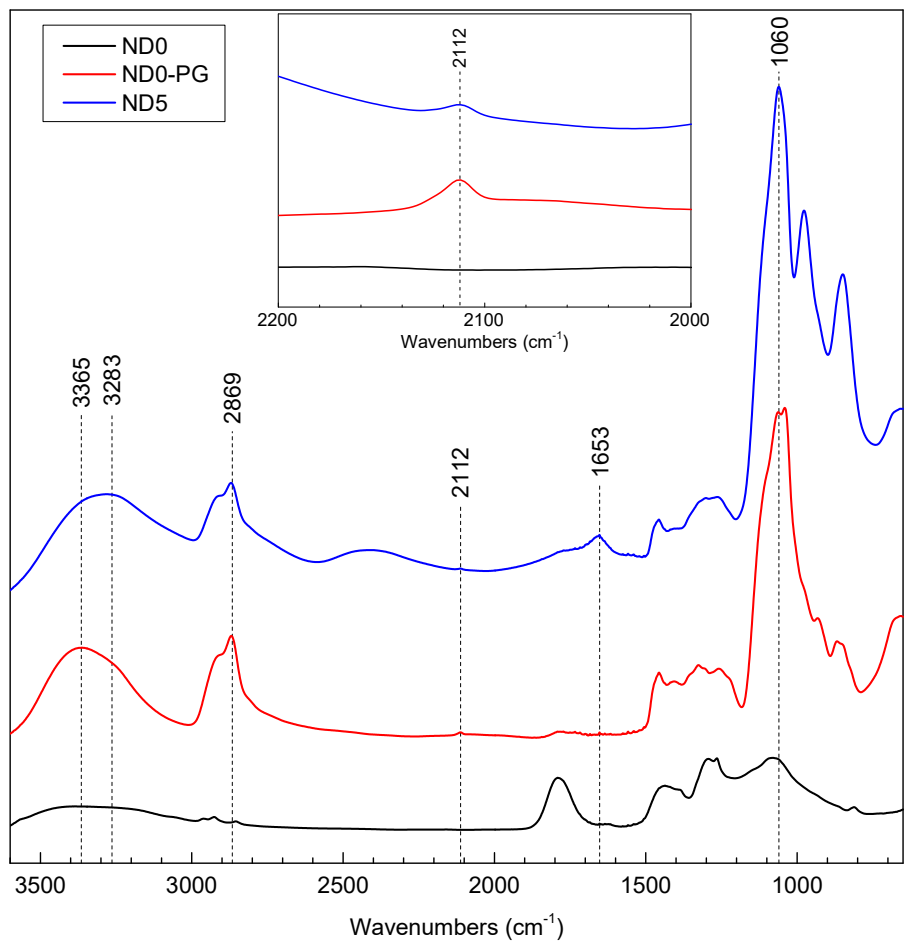

**Figure S2:** IR spectra (ATR) of **ND0** before polymer modification, **ND0-PG** with polymer, and **ND5** with nitroxide attached to the polymer. The wide and intense band at 3500-3200 corresponds to the –OH stretching band of polyglycerol and the band at 3000-2830 cm$^{-1}$ corresponds to C-H stretching band of polymer[1,2]. Also note significant changes in the fingerprint region (C-O vibrations around 1060 cm$^{-1}$). Furthermore, a weak peak at 2112 cm$^{-1}$ corresponds to alkyne in the polymer, which decreases after azide-alkyne cycloaddition (see inset for detail). The presence of TEMPO radical in the polymer is evidenced by the amide I (1653 cm$^{-1}$) band; a weaker signal of the amide II band was not observed. *Inset*: detail showing the peak at 2112 cm$^{-1}$ (magnified by a factor of 10). The lower intensity of this peak in **ND5** corresponds to the reaction of accessible alkyne groups in **ND0-PG** with azide group in **3**. All spectra are offset for clarity.

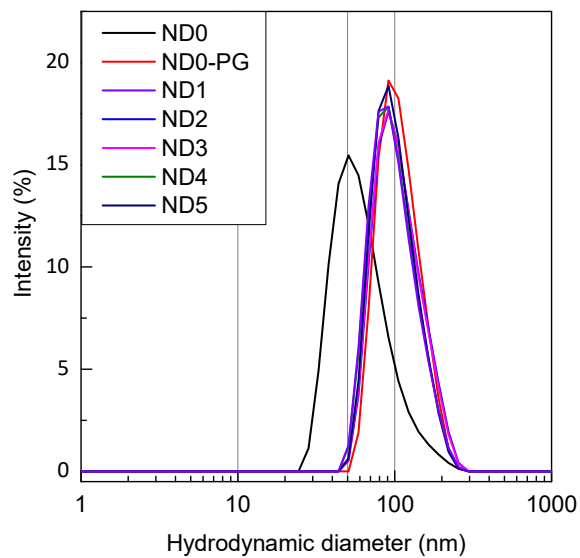

**Figure S3:** Hydrodynamic diameters of NDs analyzed by DLS. **ND0**: naked fluorescent NDs; **ND0-PG**: fluorescent NDs coated with polyglycerol-bearing propargyl moieties; **ND1-5**: **ND0-PG** particles with attached increasing amounts of the nitroxide radical. The size of the **ND0** increased by ~30 nm after polymer modification (see Table 1), while the attachment of nitroxide radicals did not cause a significant change.

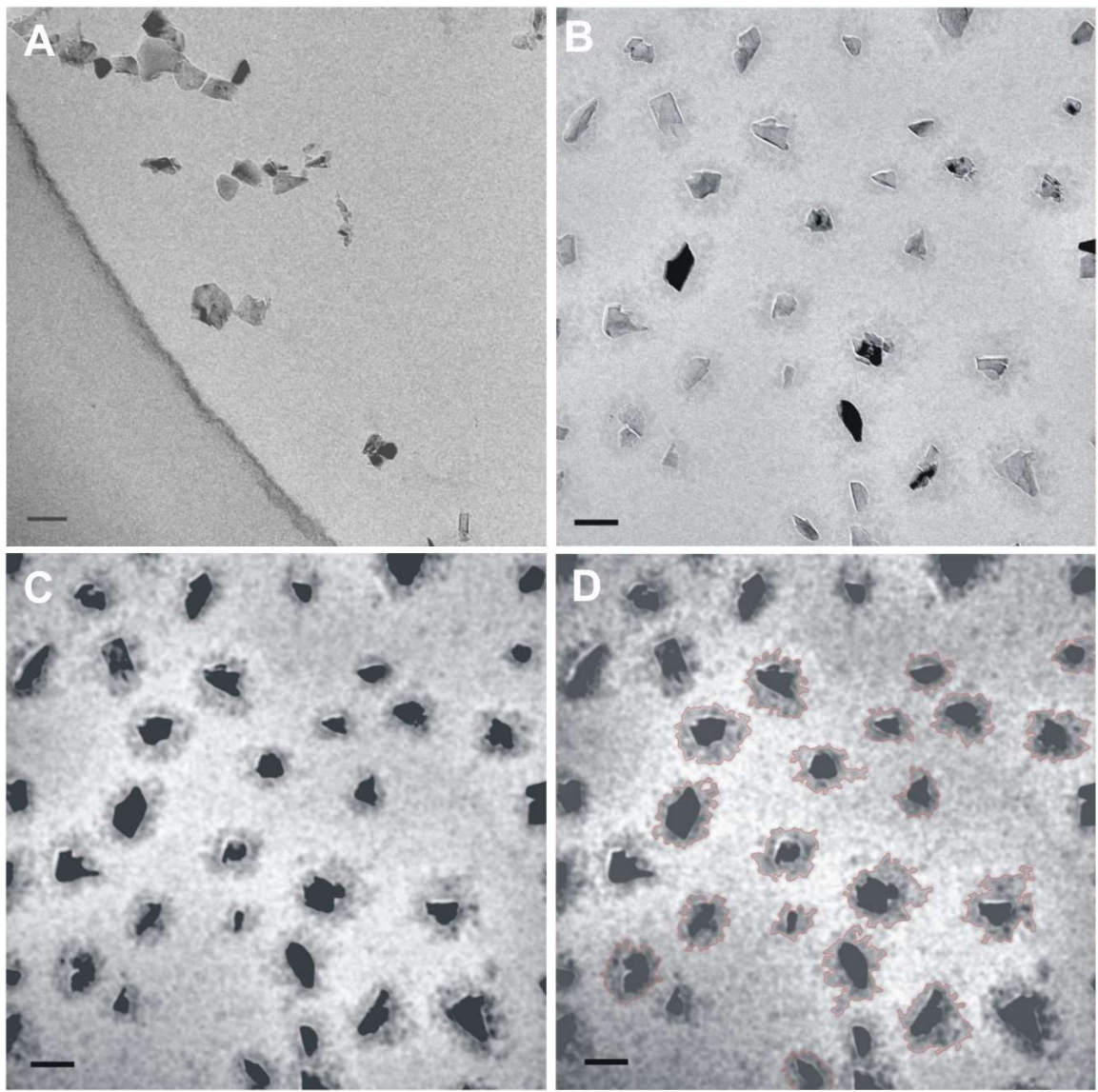

**Figure S4**: Cryo-TEM analysis of (A) naked **ND0** and (B, C, D) polymer-coated **ND5** particles used for detection of the thickness of the polymer layer on NDs. (A-B) are the original micrographs, (C) shows the image (B) processed by a band-pass filter and Gaussian blur for image analysis, (D) shows the automatically detected contours of the polyglycerol layer on selected particles in (C) using NIS software (for details, see section 1.3). Note the difference in the thickness of the polymer layer recorded *in vacuo* (Figure S5D) and in a "native" state observed by cryo-TEM (B-D). The scale bar represents 50 nm.

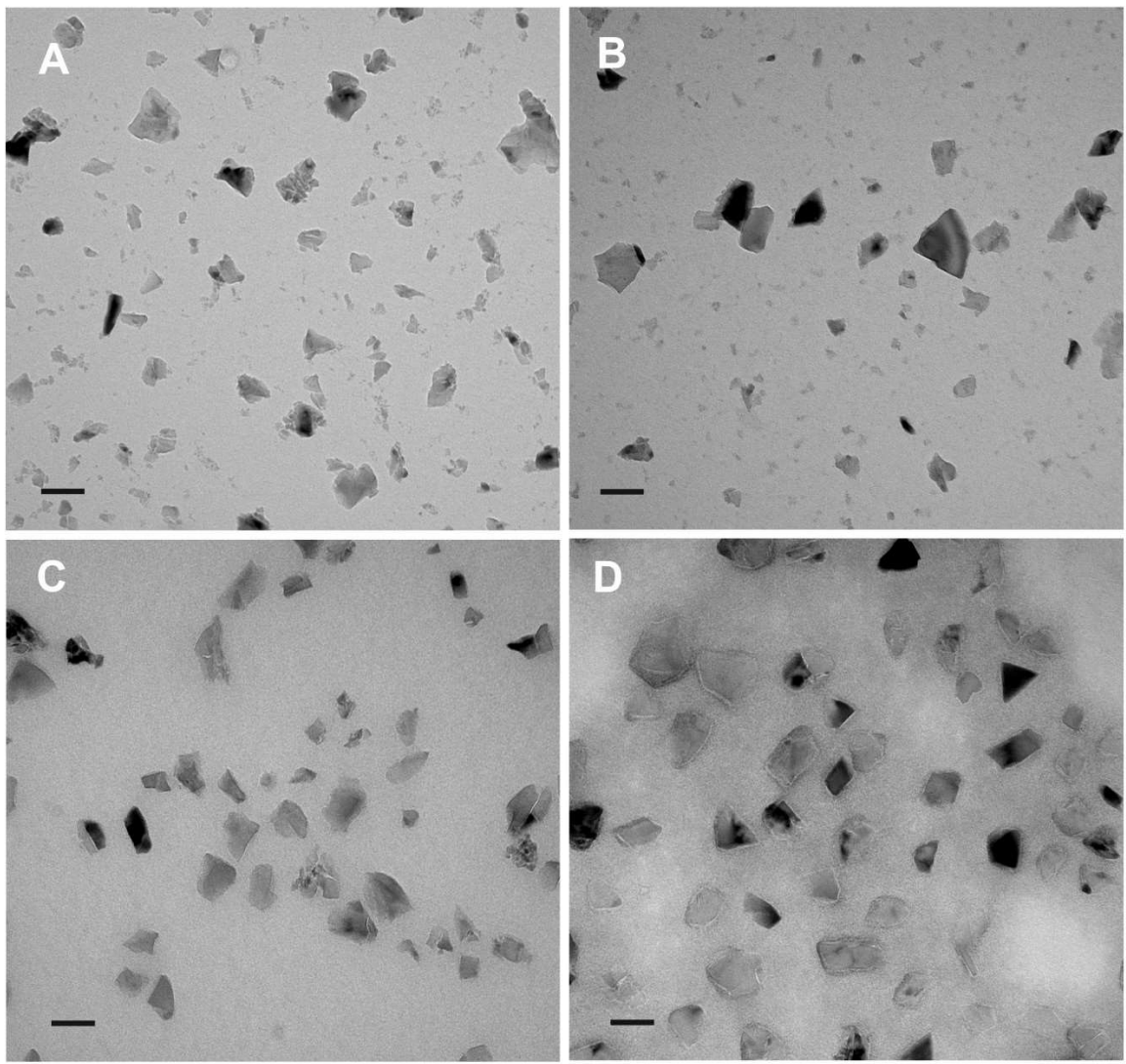

**Figure S5:** TEM images of (A, B) **ND0** and (C, D) **ND0-PG** particles. The micrographs were taken without (A, C) and with (B, D) phosphotungstic acid staining. The thin shell present on particles in (D) corresponds to a 5.7 ± 1.6 nm thick polyglycerol layer stained with phosphotungstic acid. Note the layer is collapsed due to strong dehydration caused by sample exposure to high vacuum. The scale bar represents 50 nm.

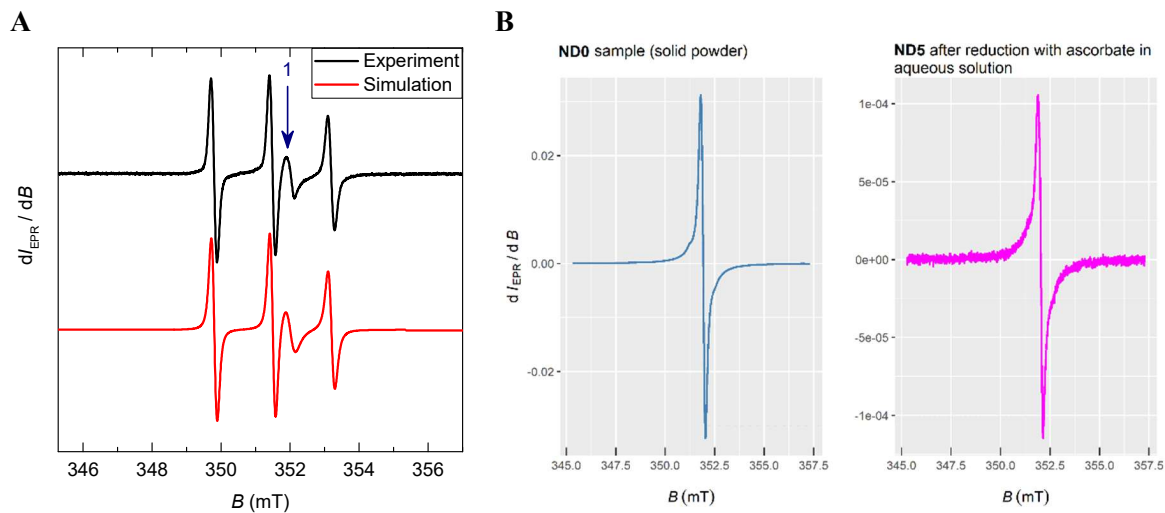

**Figure S6**: A) Simulation of nitroxide radical EPR spectrum obtained by MATLAB toolbox EasySpin[3]. The main three lines (centered at $g = 2.00555 \pm 0.00004$) represent the hyperfine splitting from the $^{14}$N nucleus. The arrow "1" shows ND "sub-surface" radicals centered at $g = 2.00254 \pm 0.00006$. B) Comparison of EPR spectra of unmodified **ND0** powder ($g = 2.00258 \pm 0.00007$; no nitroxides on the surface) and **ND5** reduced by ascorbate (no nitroxides on the surface) and measured in an aqueous environment. This signal corresponds to the one marked "1" in A). Note the missing hyperfine splitting from $^{14}$N, which is characteristic for NDs smaller than ~80 nm [4].

**Table S1**. Concentrations of nitroxide and "sub-surface" radicals in **ND1-5** determined in aqueous solutions. See section "EPR calculations" for details. To enable the comparison of all ND samples, the concentrations of radicals are expressed in $g^{-1}$ for diamond, without the polymer layer. The gradual decrease of the "sub-surface" radical concentration was most likely caused by an increased number of nitroxide radicals in the samples. The broadening of the spectral lines[5,6] caused by magnetic noise from nitroxides resulted in underestimation of the "sub-surface" radical concentration for samples with high nitroxide loads. We assume that the apparent decrease of the "sub-surface" radical concentration with increasing nitroxide load is thus an artefact of the spectral deconvolution (the concentration of sub-surface radicals should remain unchanged within the whole series of samples). The average value for **ND1-5** $(2.47 \pm 0.52) \cdot 10^{18} \, g^{-1}$ corresponds well to the value $(2.53 \pm 0.08) \cdot 10^{18} \, g^{-1}$ obtained from EPR of solid **ND0**.

| Nanodiamond | Nitroxide radicals ($g^{-1}$) | "Sub-surface" radicals ($g^{-1}$) | Nitroxide/sub-surface radicals ratio |
|---|---|---|---|
| **ND1** | 0.00E+00 | 3.15E+18 | 0.00 |
| **ND2** | 3.57E+17 | 2.80E+18 | 0.13 |
| **ND3** | 1.58E+18 | 2.62E+18 | 0.61 |
| **ND4** | 3.03E+18 | 2.09E+18 | 1.45 |
| **ND5** | 4.62E+18 | 1.68E+18 | 2.75 |

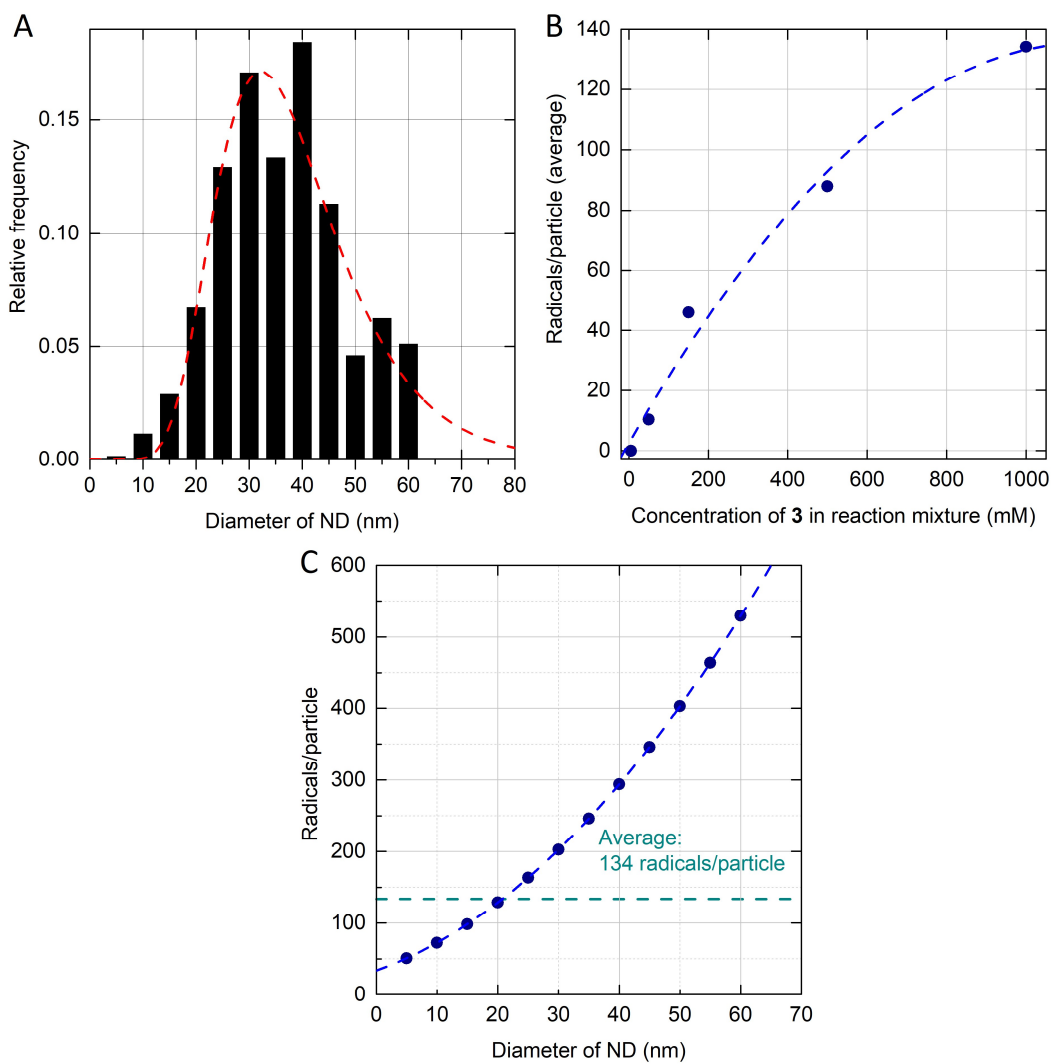

**Figure S7**: **A)** ND size distribution without the polymer layer obtained from TEM image analysis. **B)** The nitroxide load on the particles **ND1-5** as a function of the concentration of reagent **3** in the reaction mixture. After attachment, the diamagnetic **3** was converted to **2** using 20 min of sonication in cup horn in a vial exposed to air. The concentration of **2** on particles after oxidation was determined using a combination of EPR, thermogravimetry, TEM, and cryo-TEM image analysis, and NTA (for details, see section 1.3). The dashed line is shown as a guide for the eye only. **C)** Dependence of radical load on particle size for **ND5**.

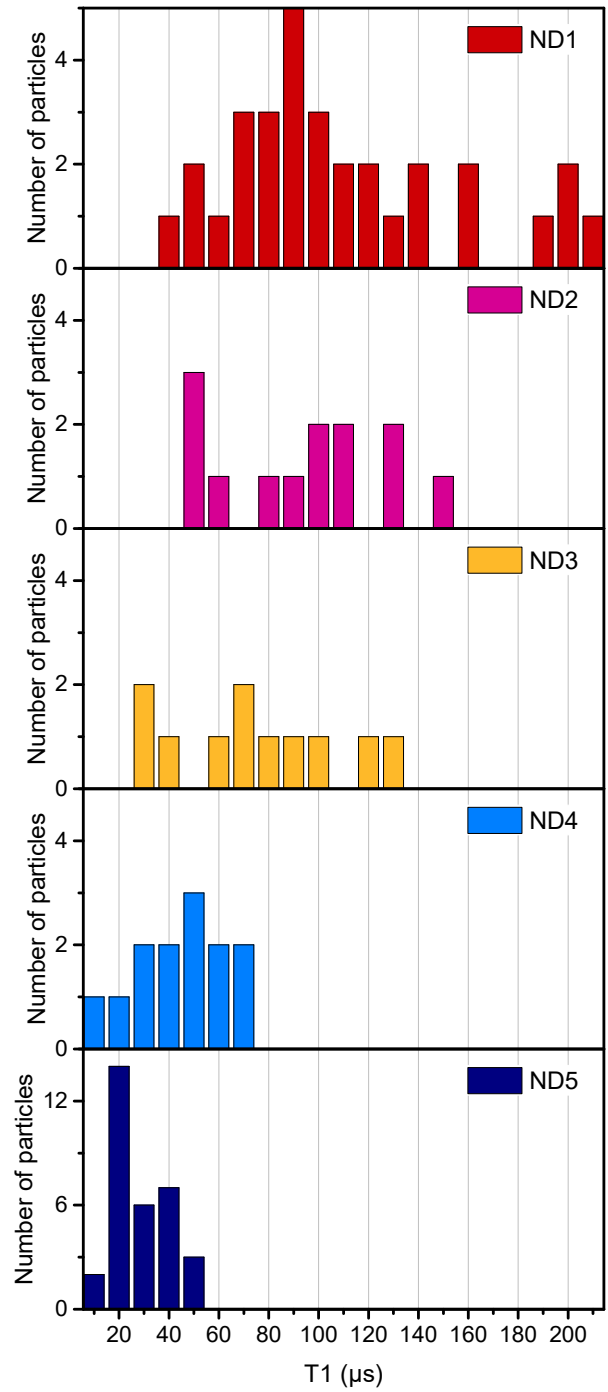

**Figure S8:** Comparison of $T_1$ histograms for samples **ND1**-**ND5**.